\documentclass{aastex631}

\usepackage{amsmath,amssymb,amsfonts}
\usepackage{graphicx}
\usepackage{subfigure}
\usepackage{fancyhdr}
\usepackage{secdot}
\usepackage{underlin}
\usepackage{natbib}

\newcommand{\di}{\partial} 
\newcommand{\e}{{\rm e}^} 
\newcommand{\apgt}{\ {\raise-.5ex\hbox{$\buildrel>\over\sim$}}\ }
\newcommand{\aplt}{\ {\raise-.5ex\hbox{$\buildrel<\over\sim$}}\ }

\def\Hcal{{\cal H}}

\def\apgt{\ {\raise-.5ex\hbox{$\buildrel>\over\sim$}}\ }
\def\aplt{\ {\raise-.5ex\hbox{$\buildrel<\over\sim$}}\ }

\shorttitle{Evolution of ice-rich bodies}
\shortauthors{Loveless et al.}

\begin{document}

\title{On the structure and long-term evolution of ice-rich bodies}

\author{Stephan Loveless} 
\affiliation{Department of Geosciences, Tel Aviv University, Tel Aviv, Israel}

\author{Dina Prialnik} 
\affiliation{Department of Geosciences, Tel Aviv University, Tel Aviv, Israel}

\author{Morris Podolak} 
\affiliation{Department of Geosciences, Tel Aviv University, Tel Aviv, Israel}

\begin{abstract}

The interest in the structure of ice-rich planetary bodies, in particular the differentiation between ice and rock, has grown due to the discovery of Kuiper belt objects and exoplanets.
We thus carry out a parameter study for a range of planetary masses $M$, yielding radii $50 \aplt R \aplt 3000$~km, and for rock/ice mass ratios between 0.25 and 4, evolving them for 4.5~Gyr in a cold environment, to obtain the present structure. We use a thermal evolution model that allows for liquid and vapor flow in a porous medium, solving mass and energy conservation equations under hydrostatic equilibrium for a spherical body in orbit around a central star. The model includes the effect of pressure on porosity and on the melting temperature, heating by long-lived radioactive isotopes, and temperature-dependent serpentinization and dehydration. We obtain the boundary in parameter space [size, rock-content] between bodies that differentiate, forming a rocky core, and those which remain undifferentiated: small bodies, bodies with a low rock content, and the largest bodies considered, which develop high internal pressures and barely attain the melting temperature. The final differentiated structure comprises a rocky core, an ice-rich mantle, and a thin dense crust below the surface.  We obtain and discuss the bulk density-radius relationship. The effect of a very cold environment is investigated and we find that at an ambient temperature of $\sim$20~K, small bodies preserve the ice in amorphous form to the present.

\end{abstract}


\section{Introduction}
\label{s:intro}

The search for extraterrestrial life has led to the pursuit of water sources, both in our own Solar System and in planetary systems around other stars. In the Solar System, except on our own planet, surface water is abundant mainly as ice, which is detected in comets, in satellites, such as Enceladus and Europa, in asteroids, such as Ceres, and in the distant planets.  Its existence is inferred---essentially from the low bulk density---for many other objects residing in the remote regions of the Solar System, such as Kuiper belt objects (KBOs) Haumea \citep{Dumas11} and Quaoar \citep{Jewitt04}, among many others. As for exoplanets, there have been more than a hundred recent discoveries of super-Earths---rocky planets with masses between 1 and 10 M$_\oplus$---orbiting other stars.  Their measured masses and radii give mean densities which indicate that many of them may be composed of mixtures of silicates and water \citep{queloz09, charbonneau09}. At present, small exoplanets are still elusive to observation, but the mere existence of large ice-rich exoplanets indicates that small ice-rich bodies may be as abundant in distant planetary systems as in our own.

The question that we address in this paper concerns the internal structure of small and intermediate-size ice-rich objects, and its evolution.
In recent years, detailed information on a few such objects has been gathered by space missions. Thus, masses, radii and moments of inertia have been obtained with great precision for Jupiter's moons  by the {\textit{Galileo}} mission and for Pluto and Charon, by {\textit{New Horizons}}. These constraints enabled the construction of models of the present interior structure of these objects \citep[e.g.,][]{Sohl02, Sotin07, McKinnon17}, assuming hydrostatic and thermal equilibrium, which match the observations with great accuracy. All these models are based on a differentiated structure---a rocky core separated from an icy mantle---regardless of prior history.  The mechanism by which such a differentiation occurs and evolves has seldom been considered in detail, perhaps because there are too many unknowns or degrees of freedom.  Although there would be very little observational evidence to confront the mechanism with, the question is interesting and worthy of investigation in a general, systematic way. Internal properties may have observable signatures that might be detected even for distant objects that are not accessible to space missions. Presently, most of such objects in our solar system populate the Kuiper belt or form icy satellites.  Although these will serve as tests to our findings, we do not aim at modeling the present structure of specific objects, which largely depends on local and present conditions. The goal is to show how differentiation may have happened and evolved. Decades ago, a mechanism was proposed by \cite{Friedson83}: upon heating, ice turns into vapor or liquid that percolate outwards, leaving behind an ice-depleted core. This mechanism, developed into an evolution code by \cite{Prialnik08}, will be pursued in the present study. 

Many studies of bodies made of ice and rock are based on solving the equilibrium structure equations for a given mass \citep[e.g.,][and the refrences therein]{leger04, Hussmann06, Noack16, VanH19b}, on dividing the evolution into separate phases \citep[e.g.,][]{Travis12}, or on following the thermal evolution and inferring structural changes based on it \citep{ORourke14, Zeng14, Bhatia17}. 
However, in order to understand the present structure of ice-rich objects, continuous self-consistent evolution models are essential.  An important advance in this respect was made by \cite{Bierson18} and \cite{Bierson19}, who calculated thermal evolution models for KBOs.  These computations were based on the assumption that the porosity of the body could be described by a time-dependent equation, as suggested by \cite{Besserer13}.  \cite{Prialnik08}, followed by \cite{Malamud13, Malamud15, Malamud16} have suggested an alternative picture, where the ice is embedded in a porous rock matrix, and the porosity changes as a result of structural and compositional changes. This model was applied to a few specific objects.     Malamud \& Prialnik also included the effect of serpentinization.  For bodies where water and rock are major components, the energy released by this process (or absorbed by its inverse) can have important consequences for thermal evolution.  

The objective of the present work is to investigate the long-term thermal history of icy bodies with radii in the range of 50-3000~km. The question we focus on is, how do ice mantles form and evolve and what are the conditions for their formation.  We thus follow the thermal and structural evolution of initially homogeneous bodies made of ice and rock over the age of the solar system (considering it typical of planetary system ages in general) by solving the time-dependent equations of heat and mass transport, taking into account all the relevant energy sources, including radioactive energy supply, phase transitions and ice-rock interactions.  The leading parameters of this study are initial size and ice content and we consider a wide range of values for both, keeping all other initial and physical parameters the same. We thus adopt a general model, applicable to the entire parameter space. We do not take into account secondary effects, such as the chemistry of ice in its different phases or the inclusion of trace volatiles within the ice, which have been studied in detail by \cite{Journaux20}, although not in the context of long-term evolution, or  loss of ice by sublimation at the surface, which was studied in detail by \cite{Malamud17}. We assume the bodies to evolve in an environment where the equilibrium temperature is sufficiently low.
In Section~\ref{s:description} we describe the model with its assumptions and input physics, in particular the equation of state (EOS); in Section~\ref{s:results} we describe and discuss the results. We conclude with a brief summary and our main conclusions in Section~\ref{s:discussion}.

\section{\bf The model}
\label{s:description}

We consider a mixture of porous rock and ice that changes with time due to energy released or absorbed in the interior by phase transitions, radioactive decay and rock-ice interactions. It is assumed that the rock forms an interconnected solid matrix and the ice is embedded in it. The rock is taken to be one single substance, thus we ignore possible chemical reactions or differentiation within the rock itself, such as separation of iron from silicates. In any case, such processes would require higher temperatures than those obtained here. We assume a "cold" start, which implies slow formation with negligible accretional heating.
As the temperature rises during evolution, the ice sublimates on the pore walls and eventually melts. The vapor and water may percolate through the rock matrix. The body is considered to be differentiated when the core becomes completely depleted of ice. The porosity decreases due to gravitational compression and almost vanishes in the interior of large bodies in the sample. Very low porosity and  shrinking pores hinder the flow of water and vapor that may cease altogether. When reaching cold outer regions, the water refreezes (and/or the vapor condenses), and an ice-rich mantle may form, where the ice/rock ratio may exceed the initial value. More details about the thermo-physical model are given in \cite{Prialnik08}. The updated input physics is described in Sections~\ref{ss:eos} -- \ref{ss:physchem}. We consider isolated hypothetical objects affected only by the host star, hence tidal heating is not taken into account. Tidal heating would require two additional free parameters, the mass of the companion and the distance between the objects, and would be relevant only in known specific cases, rather than in a parameter study. 

\newpage
\subsection{The set of equations}

The long-term evolution is computed self-consistently by the following set of equations that are solved numerically by an implicit scheme on a spherically symmetric adaptive grid, for discrete points $n$ ($n=1,\dots, N$) between the center and the surface, and time $t$. The space variable is the cumulative spherical volume $V$ and the independent variables are: the temperature $T$, the density $\rho$, and the mass fractions of all ice phases and rock---$X_s$ (solid rock), $X_a$ (amorphous ice), $X_c$ (crystalline ice), $X_v$ (water vapor)---all of which change with distance from the center and with time. Dependent variables---expressed by constitutive relations as functions of the independent ones---are the heat flux $F$, the mass fluxes $J_v$ (vapor) and $J_\ell$ (water),  specific energies denoted by $u$, the melting rate $q_\ell$ and the sublimation rate $q_v$. In this scheme, it is convenient to use integrated fluxes, namely, heat or mass crossing a spherical surface per unit time, thus the dimension of $F$ is energy per unit time, and the dimension of $J_v$ and $J_\ell$ is mass per unit time.

Amorphous ice crystallizes with a temperature-dependent timescale of $\lambda(T)^{-1}$, obtained experimentally \citep{Schmitt89}. Crystallization occurs at low temperatures, where evaporation or melting are negligible, hence the only transformation amorphous ice undergoes is crystallization (which is irreversible). 
Crystalline ice undergoes a phase transition to the vapor phase according to the local temperature (saturated vapor pressure), or to liquid phase, according to the melting temperature $T_m$, which is a function of the local hydrostatic pressure, $T_m=T_m(P)$. According to the local melting temperature, a fraction $X_\ell$ of the crystalline ice may be in liquid phase. This fraction is taken as a smoothed step function of the temperature around the melting temperature; thus, $X_\ell=X_c/[1+e^{\beta(1-T/T_m)}]$ and $q_\ell=\rho dX_\ell/dT[ \partial T/\partial t]$    \citep[see][for further details]{Prialnik08, Malamud13}. 

The equations that describe the evolution of the body are conservation equations for energy and for the masses of the various components:
\begin{eqnarray}
&&\frac{\di(\rho u)}{\di t} + \frac{\di F}{\di V}+\frac{\di(u_vJ_v + u_\ell J_\ell)}{\di V}=S \\
&&\frac{\di (\rho X_v)}{\di t} + \frac{\di J_v}{\di V}  = q_v \\
&&\frac{\di (\rho X_c)}{\di t} + \frac{\di J_\ell}{\di V}  = \lambda(T)\rho X_a - q_v\\
&&\frac{\di (\rho X_a)}{\di t} = -\lambda(T)\rho X_a,
\end{eqnarray}
where $\rho u$ denotes the sum of weighted internal energies for all species, and $S$ is energy source term, including latent heat release, the time-dependent rate of energy release by radioactive decay and the rate of energy absorption/release in the serpentinization/dehydration process. The initial conditions are total mass $M$, a homogenous composition of ice and rock in a given ratio, and an isothermal structure in hydrostatic equilibrium. The structure is assumed to evolve quasi-statically, that is, hydrostatic equilibrium is assumed to be maintained. In other words, 
conservation of momentum is replaced by the hydrostatic equation, it too expressed in terms of the space variable $V$:
\begin{equation}
4\pi \left(\frac{3V}{4\pi}\right)^{4/3}\frac{\di P}{\di V}= -\rho Gm, \qquad dm=\rho dV ,
\label{eq:hydro}
\end{equation}
satisfying the condition $\int\rho dV=M$. Formulated in this fashion, the corresponding set of difference equations is itself conservative, both in energy and in mass of ice and rock. We note that the solid matrix is not fixed, but changes with time according to changes in local composition and hence pressure. Since temporal derivatives are taken at constant $V_n$, whereas $V_n=V_n(t)$, the following transformation is implemented in the difference scheme: 
\begin{equation}
\left(\frac{\di}{\di t}\right)_V = \left(\frac{\di}{\di t}\right)_n - \left(\frac{\di V}{\di t}\right)_n 
\left(\frac{\di}{\di V}\right)_t .
\end{equation} 
The constitutive relations for the various terms and physical parameters are described in detail in \cite{Prialnik08} and \cite{Malamud15}; in the following sections we shall focus on changes made in the model, which concern the equation of state, melting temperature and serpentinization rate.

The set of differential equations is turned into a difference scheme over the grid ($n=1,\dots,N$). Together with them, we solve the grid equation required by the 
adaptive grid method, which may be symbolically expressed as
\begin{equation}
V=f(n,t) , 
\end{equation}
where $f$ is a prescribed function; here we define it so as to obtain a geometric series for the volume shells, that is, $V_{n+1}(t)-V_n(t)=q(V_n(t)-V_{n-1}(t))$, adjusting the value of $q$.

The boundary conditions at the center are zero fluxes,
\begin{equation}
F(0,t)=J_v(0,t)=J_\ell(0,t)=0
\end{equation} 
and at the surface, $R=R(N,t)$, vanishing pressure and
\begin{equation}
F(R,t)=4\pi R^2(\sigma T(R,t)^4+Z[T(R,t)]\Hcal_v-\sigma T_{\rm eq}^4),
\end{equation} 
where $Z(T)$ is the sublimation rate and $T_{\rm eq}$ is the ambient temperature (see Section~\ref{ss:grid}), assumed to be constant during evolution. We do not consider atmospheres, which may affect the boundary conditions to some extent.

The numerical scheme is implicit, based on an iterative relaxation method. A single evolution run over the age of the solar system may require up to $10^6$ time steps (time steps are determined by the code itself and change with the characteristic timescales of the evolutionary processes). In order to circumvent or overcome convergence difficulties, numerical parameters are automatically adjusted by the code along the run. This enables continuous evolution runs for 4.5~Gyr without intervention, for the entire range of models considered in this study. 

\subsection{The equation of state}
\label{ss:eos}

The crucial input physics for planetary evolution modeling is the equation of state (EOS), which relates the pressure with the density, temperature and composition. 
\begin{figure}[h]
\centering
\includegraphics[scale=0.37]{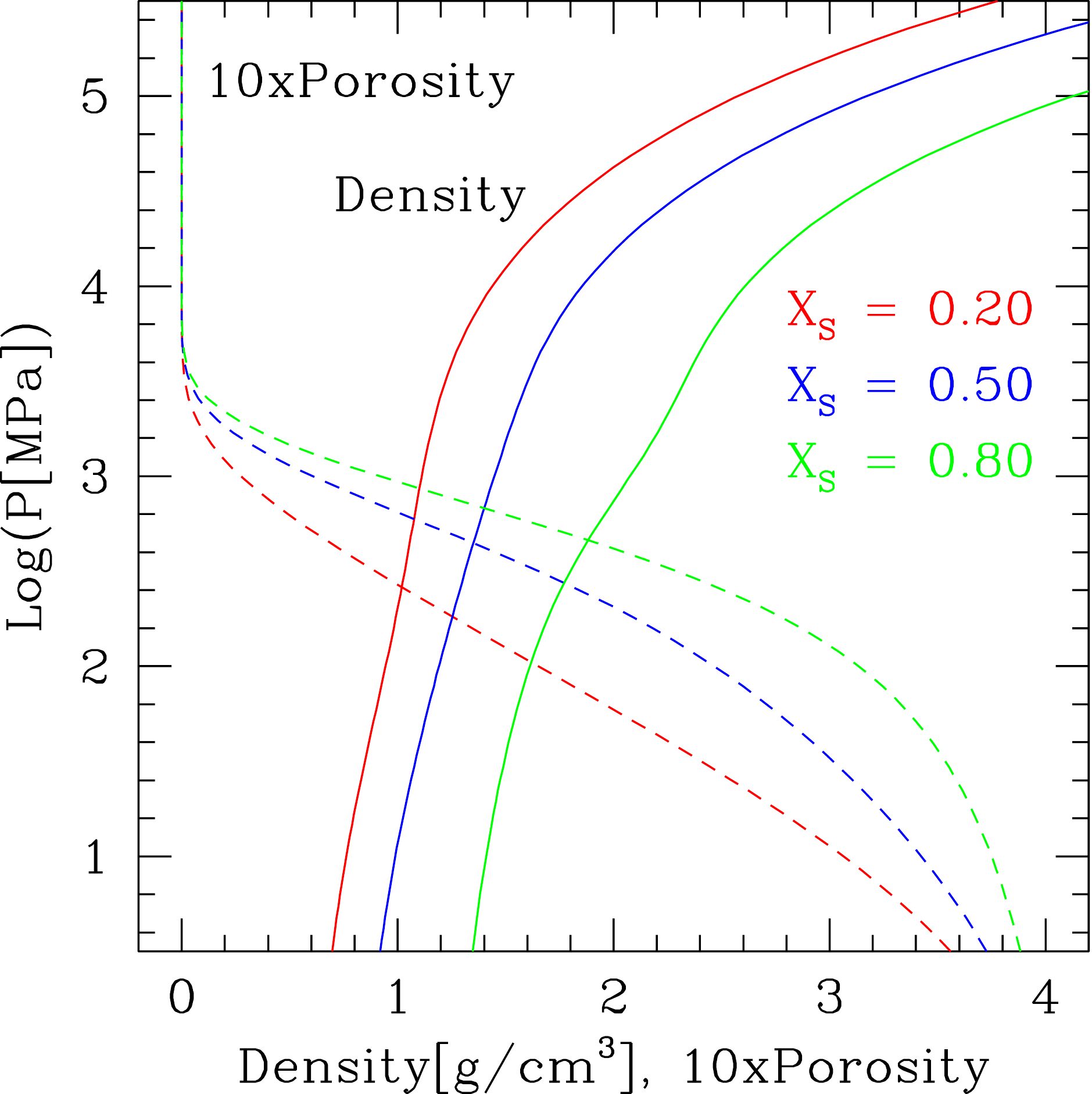}
\caption{\small{The equation of state $P(\rho,X_s)$ used in this study, shown for three different $X_s$ values. The porosity as function of pressure is also shown (multiplied by 10, for convenience). The corresponding equations are given in text. Note that the maximum value of porosity allowed is 0.4.}}
\label{fig:eost}
\end{figure}
The bodies considered are porous and therefore we have to find a solution for the pressure $P$, the bulk density $\rho$ and the porosity $\psi$ as functions of the space variable $V$. We assume a mixture of porous ice and porous rock, where the mass fraction of rock is $X_s$ and the mass fraction of ice is $X_i=1-X_s$. The porosity of the ice is denoted by $\psi_i$ and that of the rock, by $\psi_s$; the densities of nonporous ice and rock are denoted by $\bar\rho_i$ and $\bar\rho_s$, respectively. 
For a given pressure $P$ and ice mass fraction $X_i$, the bulk density weighted by the volume fractions of the components is thus given by
\begin{equation}
\label{eq:density1}
\frac{1}{\rho}= \frac{X_i}{\bar\rho_i(P)[1-\psi_i(P)]}+\frac{1-X_i}{\bar\rho_s(P)[1-\psi_s(P)]}.
\end{equation}
We adopt the second-order Birch-Murnaghan (BM) approximation for the ice and rock densities, where the coefficients are chosen so as to achieve a close agreement with the tabulated EOS developed and used by e.g., \cite{vazan13, vazan15}, based on the quotidian EOS (QEOS) of \cite{more88}. Hence $\bar\rho_i$ is obtained by solving
\begin{equation}
\label{eq:eosi}
P_{0,i}\left[\left(\frac{\bar\rho_i(P)}{\rho_{0,i}}\right)^{7/3}-\left(\frac{\bar\rho_i(P)}{\rho_{0,i}}\right)^{5/3}\right]-P =0,
\end{equation}
where $\rho_{0,i}=0.917$~g/cm$^3$ and $P_{0,i}=2.59\times10^{11}$~dyn/cm$^2$, and similarly, $\bar\rho_s$ is obtained by solving
\begin{equation}
P_{0,s}\left[\left(\frac{\bar\rho_s(P)}{\rho_{0,s}}\right)^{7/3}-\left(\frac{\bar\rho_s(P)}{\rho_{0,s}}\right)^{5/3}\right]-P=0,
\label{eq:eosd}
\end{equation}
where $\rho_{0,s}=3.25$~g/cm$^3$ and $P_{0,s}=1.28\times10^{12}$~dyn/cm$^2$.
In the range of interest for densities and temperatures, the effect of temperature on pressure according to the QEOS is very small. We have also estimated the Debye correction to the BM EOS and found it small. 
Therefore, we assume the pressure to be independent of temperature, which is valid for the entire relevant temperature range.
The porosity, which allows the flow of vapor and water through the rock, thus enabling the differentiation, is an important feature of the model. The mutual effect of pressure on porosity is taken into account empirically, 
\begin{equation}
\label{eq:pori}
\psi_i(P)=0.45\exp(-4.7434\times 10^{-5}\sqrt{P})
\end{equation}
\begin{equation}
\label{eq:pord}
\psi_s(P)=0.4\exp(-1.28\times 10^{-10}P),
\end{equation}
where $P$ is given in dyn/cm$^2$ (see \cite{Malamud15} and references therein),
but the effect of temperature on porosity is not. While, clearly, an increasing pressure will reduce porosity, the effect of temperature on porosity is more controversial and depends critically on composition and rock structure. Although an increasing temperature favors compaction \citep{Neumann14}, it may also weaken the structure and open cracks, thus increasing the effective porosity \citep{Qi21}. 

Taking the derivative of eq.(\ref{eq:hydro}) with respect to $V$, we obtain the hydrostatic equation in the form
\begin{equation} 
\label{eq:hydro1}
\frac{d}{dV}\left(\frac{V^{4/3}}{\rho}\frac{dP}{d\rho}\frac{d\rho}{dV}\right)=-\left(\frac{4\pi}{81}\right)^{1/3}G\rho ,
\end{equation}
which requires the functional dependence $P(\rho)$ in order to obtain the bulk density profile. Hence, substituting the four functions of $P$ [eqs.~(\ref{eq:eosi})-(\ref{eq:pord})] in eq.(\ref{eq:density1}), we solve it implicitly to obtain the functional dependence of $P$  and  $dP/d\rho$ on $\rho$, to be substituted in eq.~(\ref{eq:hydro1}).
Finally, the bulk local porosity is given by
\begin{equation}
\label{eq:porosity}
\psi=\frac{\frac{X_i\psi_i}{1-\psi_i}+\frac{(1-X_i)\psi_s}{1-\psi_s}}{\frac{X_i}{1-\psi_i}+\frac{1-X_i}{1-\psi_s}}.
\end{equation}
This approach is different from that considered by \cite{Malamud15}, which was used for mid-sized bodies---up to a few hundred km in radius---but is not well suited to the high pressures that arise in larger (more massive) bodies. The aim of the present study is to detect trends of behavior with changing mass (size), and hence it is important to apply the same physical model and assumptions to all cases. The EOS developed and used here is applicable to a much wider range of sizes; 
it is shown in Fig.~\ref{fig:eost}. 

\subsection{The melting temperature}
\label{ss:melt}

\begin{figure}[h!]
\centering
\includegraphics[scale=0.35]{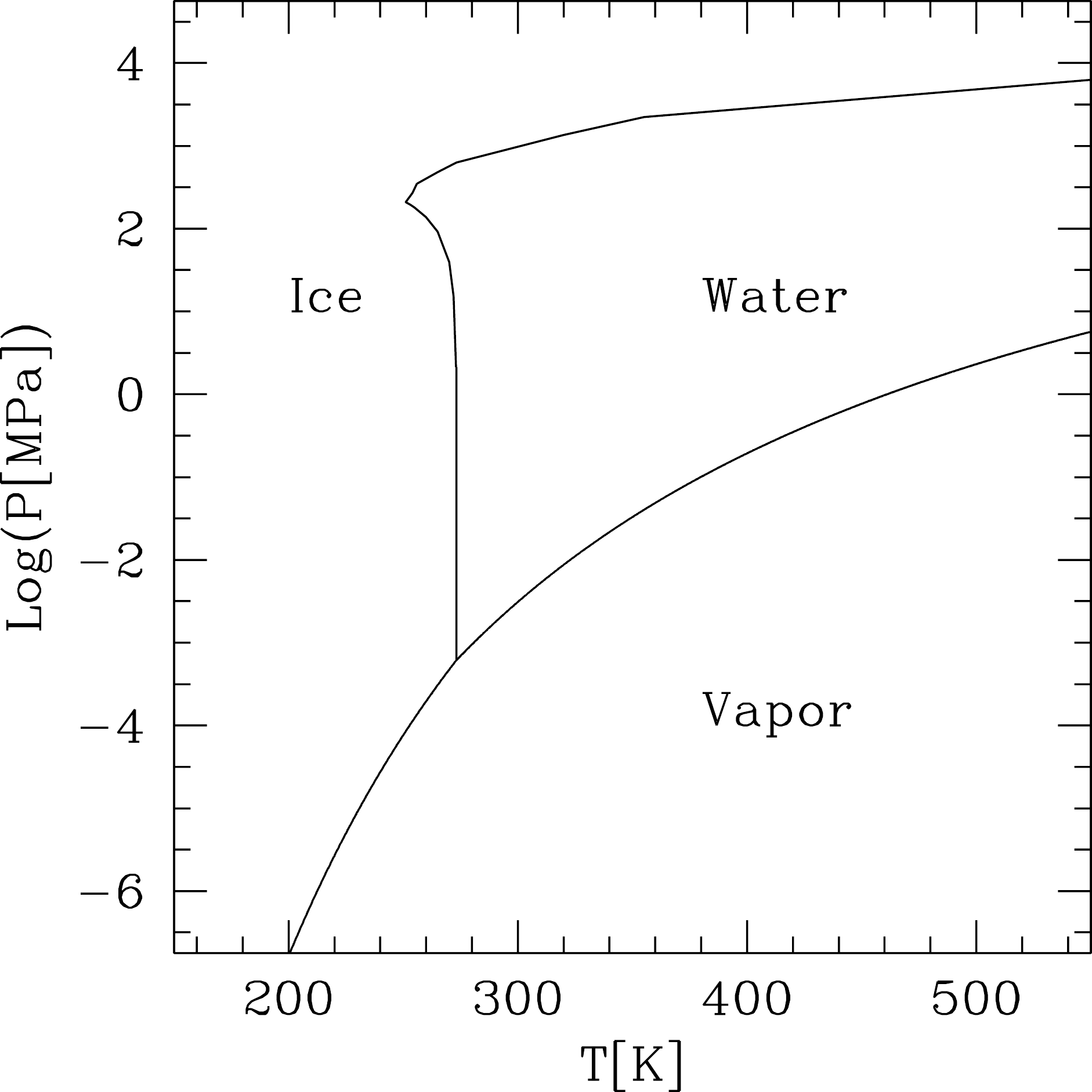}
\caption{\small{Simplified phase diagram used in this study for the relevant ranges of ice temperature and pressure. Transitions between ice phases at high pressures are ignored. The transition between amorphous and crystalline ice is not shown.}}
\label{fig:phased}
\end{figure}

The melting temperature $T_m$ of ice is constant at 273.16~K up to a few MPa; at higher pressures, it first drops to 250~K and then rises sharply with pressure up to and beyond 500~K  \citep{Choukroun07}. This is shown in the simplified phase diagram used in this study, Fig.~\ref{fig:phased}. 

The effect of anti-freeze compounds, such as ammonia, that lower the melting temperature  \citep{Leliwa02} has been neglected. To track this effect would have required the addition of another free parameter, since the abundance of anti-freeze substances is not well-known and may vary among different objects. In the presence of anti-freeze, core formation may start earlier.

\subsection{Serpentinization/dehydration}
\label{ss:physchem}

Low temperatures will always prevail in the outer part of an ice-rich planetary object, while the central part is heated. At some point, between the center and the surface, temperatures will be in the proper range for liquid water and hence interaction between water and rock is expected \citep[see][]{Malamud16}. Pristine silicate rock interacts with water in serpentinization reactions, which are exothermic chemical reactions. Here we adopt the serpentinization rate given by \cite{ruepke17} in units of s$^{-1}$,
\begin{equation}
f_s(T)=A_s\exp(-b/T)\left[1-\exp(-c(1/T-1/T_0))\right],
\end{equation}
where $A_s=808.3\times10^{-13}$~s$^{-1}$, $b=3640$~K $T_0=623.6$~K and $c=8759$~K. This function peaks around 300~K; the energy released is $2.5\times10^5$~J/kg.
Serpentinization changes the ice/rock ratio and generates substantial amounts of energy, comparable to the latent heat of ice melting. A reaction runaway is therefore possible \citep{Jewitt07}, depending on the local thermal conductivity, where the serpentinization front advances through the medium feeding on its own energy. It also lowers the specific density of the rock by up to 20\%, but this effect is not included in our EOS. We have experimented with a 10\% lower value of the specific rock density and found the effect to be negligible. 

The inverse process to serpentinization is rock dehydration. Once the rock is dehydrated, water is released back into the body, while the rock mass is reduced by the same amount. The rate of serpentine dehydration is a function of temperature, and normally high temperatures are required. Here we adopt the rate derived by \cite{Sawai13} in the form
\begin{equation}
f_h(T)=A_h\exp(-h/T),
\end{equation}
where $A_h=3.2\times10^{-3}$~s$^{-1}$ and $h=26354$~K. The reaction becomes important at $\sim$700~K. The energy absorbed is of order $4\times10^5$J/kg. At 700~K, this amounts to almost half of the specific energy of the rock, hence the cooling effect will be strong and the reaction is thus expected to occur at much higher temperatures. We note that if water is first absorbed and later released from the rock, a net loss of energy results.

\subsection{The grid of models, spanning the parameter space}
\label{ss:grid}

Observations of distant objects and of exoplanets provide limited information regarding their structure, hence we do not attempt the simulation of particular objects, but rather carry out a systematic parameter study, looking for correlations between various observable properties (mass, radius, composition), and between such properties and the internal configuration of the objects.
Simple $R(M)$ relationships have been obtained from a variety of models \citep[e.g.,][]{Seager07,Howe15,Zeng16} and the internal structure has been studied by, e.g., \cite{Spiegel14}, but not as the outcome of long-term evolution. Rather, as mentioned in Section~\ref{s:intro}, a differentiated structure is generally assumed at the beginning, with the core size as a free (adjustable) parameter. 

In the present study we start with a homogeneous configuration; differentiation will eventually result from thermal and structural evolution.
A comprehensive study of ice-rich bodies faces the problem of a large number of free parameters, both regarding the object itself, such as composition and structural properties, and regarding its environment, such as the distance from the central star and the luminosity of the star.  We thus focus on two independent and observationally relevant parameters and compute models for all possible (and viable) combinations of these parameters: mass and ice/rock ratio.  
\begin{deluxetable*}{lccccc}[h!]
\tablenum{1}
\tablecaption{Initial model masses (kg) grouped by the rock mass fraction ($X_s$)}
\tablewidth{0pt}
\tablehead{
\colhead{Size bin (km)} & \colhead{$X_s=0.20$} &  \colhead{$X_s=0.35$} & \colhead{$X_s=0.50$} & \colhead{$X_s=0.65$} &  \colhead{$X_s=0.80$}
}
\label{tab:masses}
\startdata
{50} &$6.28\times10^{17}$&$7.44\times10^{17}$&$7.44\times10^{17}$&$7.33\times10^{17}$&$9.95\times10^{17}$\\
{100}&$3.77\times10^{18}$&$3.98\times10^{18}$&$3.98\times10^{18}$&$4.40\times10^{18}$&$5.45\times10^{18}$\\
{300}&$1.07\times10^{20}$&$1.13\times10^{20}$&$1.31\times10^{20}$&$1.24\times10^{20}$&$1.58\times10^{20}$\\
{600}&$9.95\times10^{20}$&$1.09\times10^{21}$&$1.09\times10^{21}$&$1.09\times10^{21}$&$1.36\times10^{21}$\\
{1200} &$6.88\times10^{21}$&$7.60\times10^{21}$&$8.69\times10^{21}$&$1.09\times10^{22}$&$1.23\times10^{22}$\\
{1800} &$2.44\times10^{22}$&$2.93\times10^{22}$&$3.24\times10^{22}$&$3.42\times10^{22}$&$4.64\times10^{22}$\\
{2400} &$5.79\times10^{22}$&$6.95\times10^{22}$&$8.11\times10^{22}$&$9.27\times10^{22}$&$1.19\times10^{23}$\\
{3000} &$1.19\times10^{23}$&$1.47\times10^{23}$&$1.70\times10^{23}$&$2.04\times10^{23}$&$2.43\times10^{23}$\\
\enddata
\end{deluxetable*}

We consider masses in the range $6\times10^{17}-2\times10^{23}$~kg, which yield final (present-day) radii roughly between 50--3000~km, noting that radii are determined by the hydrostatic equation and change with time. Since it is more customary to identify observed objects by their radii, the models will be referred to in what follows by their final radii, rather than by their masses. The models may be divided into 8 different size bins, as listed in Table~\ref{tab:masses}; as radii change during evolution, we choose the initial masses such that the final radii fall within 20\% of the bin radius.

\begin{figure}[h!]
\centering
\includegraphics{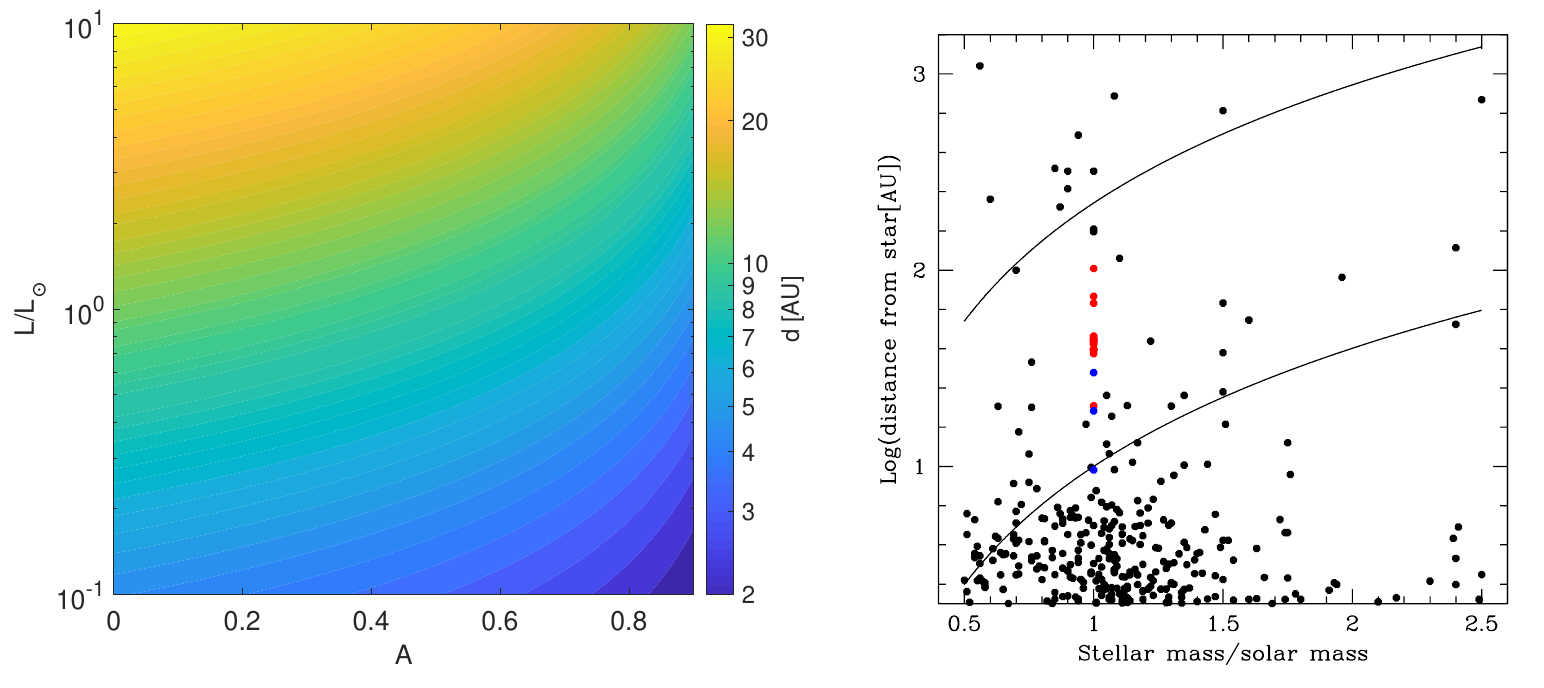}
\caption{\small{Left: Region of parameter space [$L,A,d$] corresponding to $T_{\rm eq}=90$~K. Right: Distance from central star as function of the stellar mass where the ambient temperature is 90~K (lower curve) and 20~K (upper curve). Exoplanets are marked, data from the NASA Exoplanet Archive http://exoplanetarchive.ipac.caltech.edu, as well as KBOs (red) and satellites (blue).}}
\label{fig:bound}
\end{figure}

We consider 5 different ice-rock mixtures, for: $X_s$=0.8, 0.65, 0.5, 0.35 and 0.2, hence ice/rock ratios between 0.25 and 4. A composition dominated by ice may not be well described by a rocky matrix with ice embedded in it. Such a composition is more of academic interest than encountered in reality. Even comets, which are the most ice-rich objects in our solar system, appear to have quite low ice/rock ratios \citep{Rotundi15}. We consider the high ratios in order to discern trends. The parameter space is thus represented by 40 models. There are three additional parameters that must be determined: the albedo $A$, the stellar luminosity $L$ and the distance from the star $d$. These may be combined into a single free parameter that we refer to as ambient temperature, and we define by $\sigma T_{\rm eq}^4=(1-A)L/16\pi d^2$.  We adopt as baseline an ambient equilibrium temperature of 90~K---sufficiently low for sublimation at the surface to be negligible (so that mass and bulk ice/rock ratios are conserved)---and 20~K as a far extreme (for a few cases). In Fig.~\ref{fig:bound}-left we show the parameter combinations for which our choice of $T_{\rm eq}=90$ apply. For planets around main-sequence stars, the chosen values of  $T_{\rm eq}$ correspond to limiting curves in the stellar mass -- distance plane as shown in Fig~\ref{fig:bound}-right. Also marked in the figure are exoplanets; those relevant to our study lie between the two curves.

\subsection{Initial and physical parameters}

The initial and physical parameters used are listed in Table  \ref{tab:init}. We start with fully formed objects of given mass. We do not take into account the effect of short-lived radioactive nuclei (such as $^{26}$Al or $^{60}$Fe), which may be significant during the accretion phase of the objects if the two processes have similar timescales (a few Myr \citep{Merk06}). As the main effect of short-lived radionuclides on early evolution would be of heating the interior sufficiently for the ice to crystallize, we start with crystalline ice for the $T_{\rm eq}=90$~K. In a colder and hence more distant environment, the formation times are much longer than a few Myr, hence the effect of short-lived radionuclides should be negligible. But for these models, it is reasonable to assume the ice to be initially amorphous. We do not take into account the gravitational energy released in restructuring, which is negligibly small compared with the other heat sources \citep{Malamud15}. 

\begin{deluxetable*}{ll}
\tablenum{2}
\tablecaption{Initial and physical parameters}
\tablewidth{0pt}
\tablehead{
\colhead{Parameter} &  \colhead{Value} 
}
\label{tab:init}
\startdata
Initial uniform temperature            &  $T_{\rm eq}$ \\
Initial $^{40}$K abundance           &  $1.13\times10^{-6}$~ppm\\
Initial $^{232}$Th abundance      &  $5.52\times 10^{-8}$~ppm\\
Initial $^{235}$U abundance          &  $6.16\times 10^{-9}$~ppm \\
Initial $^{238}$U abundance         &  $2.18\times 10^{-8}$~ppm \\
Pore size  range                            &   0.1$\mu$m---1~cm \\
Ice specific energy                        &  $3.75\times 10^4 T^2 + 9.0\times 10^5 T$~erg~g$^{-1}$\\
Water specific energy                   &  $4.187\times 10^7$~erg~g$^{-1}$\\
Rock specific energy                    &   $1.3\times 10^7$~erg~g$^{-1}$\\
Water thermal conductivity           & $5.5\times 10^4$~erg~cm$^{-1}$~s$^{-1}$~K$^{-1}$\\
Ice thermal conductivity (c)           &  $5.67\times 10^7/T$~erg~cm$^{-1}$~s$^{-1}$~K$^{-1}$ \\ 
Ice thermal conductivity (a)          &  $2.348\times 10^2T+2.82\times 10^3$~erg~cm$^{-1}$~s$^{-1}$~K$^{-1}$ \\
Rock thermal conductivity            &  $2\times 10^5$~erg~cm$^{-1}$~s$^{-1}$~K$^{-1}$\\
Water dynamic viscosity               & $2.939\times10^{-4}\exp{\left(\frac{507.88}{T-149.3}\right)}$~dyn~cm$^{-2}$~s\\
                                                     &          [ $5.05\exp{(-5.71T/T_m)}$ for high $T_m$]\\
Latent heat of crystallization         &  $9\times 10^8$~erg~g$^{-1}$\\
Latent heat of melting                   & $3.34\times 10^9$~erg~g$^{-1}$\\
Latent heat of sublimation           &   $A_0- A_1T+A_2T^2-A_3T^3$~erg~g$^{-1}$, where\\
 & $A_0=3.714\times10^{10},\ A_1=7.823\times10^7,$\\
 &$A_2=1.761\times10^5,\ A_3=1.902\times10^2$\\
 Crystallization rate                 & $1.05\times 10^{13}\exp{-5370/T}$~s$^{-1}$ \\
\enddata 
\tablecomments{The thermal conductivity is corrected for porosity by a factor $(1-\psi^{2/3})$ and pore sizes shrink when the porosity decreases (therefore the range of sizes).}
\end{deluxetable*}

The assumed initial temperature as the local equilibrium temperature has no effect (so long as it is below melting temperature) since, as we shall see, evolution soon erases initial conditions. We thus assume a "cold start" (as did \cite{Bierson19} for the evolution of KBOs), but we would like to point out the argument of \cite{Bierson20} that Pluto's extensional tectonics and lack of compressional features would be more consistent with a "hot start".  The initial conditions for such a hot start, while having an influence on the subsequent evolution of the body, are poorly constrained, and a proper study is beyond the scope of the present work. \cite{Monteux14} have shown that rapid accretion of minor planets by large impactors---a few to a few hundred km in size---would impart sufficient energy to an accreting body of 1000~km or more,  to induce melting and differentiation already upon formation. The assumption of a low initial temperature implies a relatively slow accretion rate by small impactors and is compatible, for example, with \cite{Kenyon12} or with the pebble accretion scenario \citep{Chiang10,Morbidelli20}. 

\section{Results of evolutionary calculations}
\label{s:results}

The course of evolution varies considerably both with the mass of an object and with its initial composition. The common feature is a rise in temperature due to radioactive heating. While small objects reach a maximum temperature that subsequently starts to decline towards the local equilibrium value, the larger objects reach and maintain high temperatures even to the present day. Objects with a lower rock content, and hence proportionally lower abundances of radioactive species, reach lower temperatures than those of similar size with a higher rock content. 

As many different factors act in tandem or compete with each other, it is difficult to predict the outcome of long-term evolution intuitively, and sometimes the variation of properties with initial parameters is not monotonic as might have been expected. For example, local heating occurs due to radioactive decay, serpentinization, condensation, crystallization (when relevant), conduction and advection from hotter regions, each of which depends on composition and on local structure, such as density or porosity. Local cooling occurs due to dehydration, sublimation/evaporation in pores, as well as conduction and advection towards cooler regions. In addition, the melting temperature at high pressures is not monotonic with pressure; it goes through a minimum and then increases sharply. In the following sections we shall show and discuss the results from different angles.

\subsection{Evolution of characteristic properties}
\label{ss:evol}

\begin{figure}[t]
\centering
\includegraphics[scale=0.30]{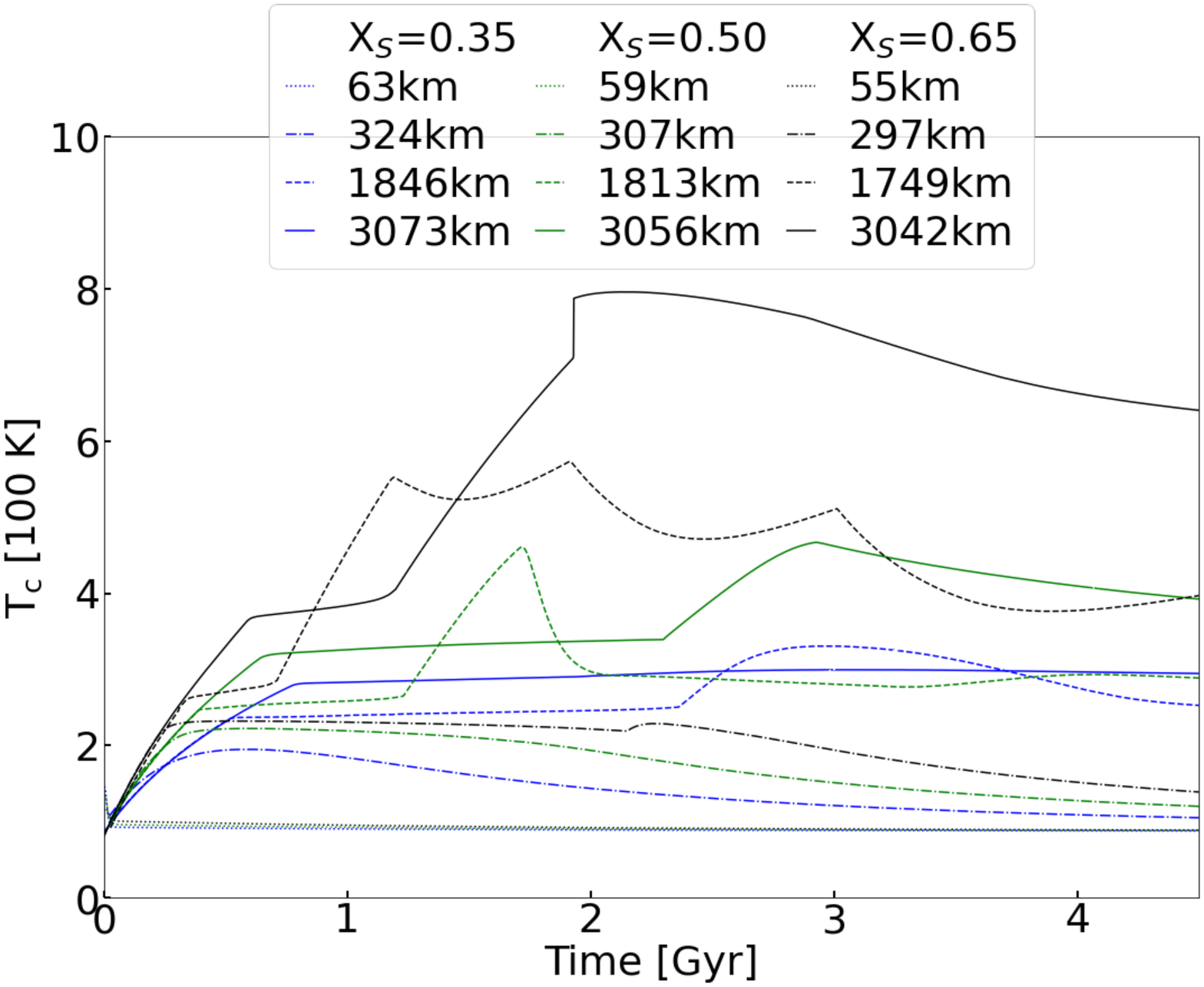}
\caption{\small{The evolution of the central temperature for models of different initial masses and compositions, identified by their final radii. A plateau marks ice melting, which occurs earlier and proceeds faster in rock-rich models. Rock-rich models also attain higher temperatures because the amount of radioactive species is correspondingly higher.}}
\label{fig:Tc-evol}
\end{figure}

The temperature is the driver of all processes and since the central temperature gives a good representation of the internal temperature regime, we show in Fig.~\ref{fig:Tc-evol}  the evolutionary course of the central temperature for a few illustrative cases. The long-term internal energy source decays exponentially, hence the temperature is expected to rise at the beginning, reach a peak and then decline. The rise and decline are, however, affected by the thermal time scales that are size-dependent, hence the evolutionary outcome varies. The larger bodies reach 4.5~Gyr while still heating up or just starting to cool, while  steady state -- equilibrium with the environment -- is reached only by very small objects. 

If and when the melting temperature is attained, 
the liquid water diffuses from the hot inner regions outwards and a rocky core forms. This only occurs in objects larger than $\sim300$~km. The water refreezes when it reaches cold outer layers, forming an ice-rich mantle. 
Smaller bodies do not reach melting temperature nor high enough temperatures for sublimation to be significant before they start cooling, and thus remain homogeneous. 

Migration of water from hot to cold regions starts even before melting temperature is attained as a result of evaporation in pores. The effect of vapor, which fills the pores and may flow through the body, is twofold: first, it contributes to water depletion; secondly, it contributes to heat transfer, both by advection and by absorption of latent heat by sublimation in hot regions and release of latent heat by recondensation in colder parts of the body. 
The plateau in the central temperature evolution shown in Fig.~\ref{fig:Tc-evol} corresponds to ice melting, which proceeds rapidly for high $X_s$ (rock content) objects, since the amount of radioactive species is proportional to the amount of rock. The plateau value increases with size of the object, as the central pressure increases and with it, the melting temperature. 
If the temperature rises above $\sim350$~K in regions where water is present, serpentinization occurs and the water reacts with the rock and is absorbed by it. 
Since the process releases energy, the local temperature rises and the process speeds up. This effect is stronger in objects that reach higher temperatures, as the rate of serpentinization is strongly temperature-dependent. It stops when water is depleted or when the rock becomes saturated. If still higher temperatures are attained ($\sim700$~K), the reverse reaction takes place, the rock undergoes dehydration and the exuded water flows rapidly towards colder regions. Heat is absorbed in the process, which causes a temporary decline in temperature, after which the supply of radioactive energy resumes control. The oscillations in temperature illustrated  in Fig.~\ref{fig:Tc-evol} are due in part to the ice-rock interactions, and in part to freezing and melting. 

\begin{figure}[h!]
\centering
\includegraphics{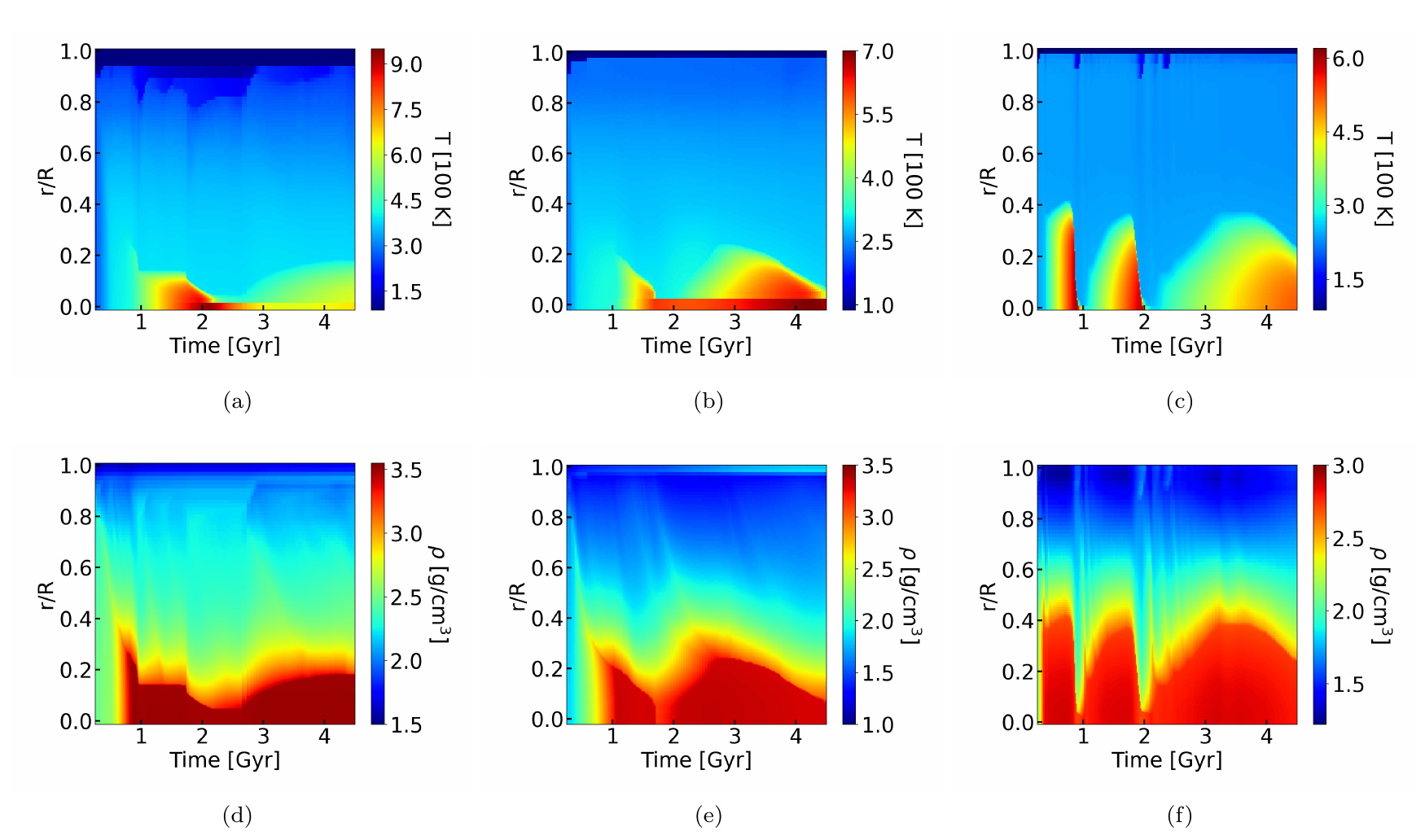}
\caption{\small{Evolution of the temperature profile (top) and the density profile (bottom) with respect to normalized radius $r/R$ for three combinations of initial parameters: 3000~km and $X_s=0.80$ (left), 2400~km and $X_s=0.65$ (middle), 1200~km and $X_s=0.8$ (right) for 4.5 Gyr. Note that due small changes in the outer radius, $r/R$ may be slightly shifted with respect to $r$.}}
\label{fig:coreform}
\end{figure}

Examples of the differentiation process are shown in Fig.~\ref{fig:coreform}. On the evolution time scale it starts rapidly, once the core temperature reaches the melting point of ice, just before 1~Gyr. 
An extended rocky core forms temporarily when the water flows outwards and refreezes or is absorbed by the rock. The ice-depleted core contracts, the pressure rises and with it, the melting temperature. When eventually, the region above the core reaches the melting point, the water diffuses back inward in part, and refreezes there, because the melting temperature is higher than in its place of origin. As the core continues to heat up, if the temperatures rise sufficiently, melting and ice depletion occur again. In larger bodies, where the core pressures are high, the melting temperature is high as well and this process is weak (left panels in Fig.~\ref{fig:coreform}). With decreasing mass, the process is stronger and repeats a few times (right panels in Fig.~\ref{fig:coreform}). In still smaller objects, where internal pressures are lower, the melting temperature is less sensitive to the pressure, and these oscillations disappear. Thus, the evolution of core formation is driven by the competition between rising temperature and rising pressure and hence melting temperature.

The heat flux crossing the surface is much smaller than the reflected and reemitted (as thermal radiation) stellar energy. Hence the stellar energy serves mainly to determine the surface temperature, which is always within a few degrees of the local equilibrium temperature  (see also \cite{VanH19a, VanH19b} ). Nevertheless, even a difference of a few degrees is capable of affecting the thermal budget.
Only a negligible amount of water ice is lost by vapor flow at the surface, no more than $\sim 10^{-5}$ of the mass, hence the ice/rock ratio is not affected by it. Nevertheless, this may indicate the possible formation of an atmosphere. 
 
Due to the internal composition changes, the physical structure changes as well, as illustrated in the lower panels of Fig.~\ref{fig:coreform} by the changes of the bulk density throughout evolution. This may cause stresses that would show on the surface as fissures or cracks, such as have been detected on Pluto and on icy satellites \citep[e.g.,][]{Spencer20}. 
The model on the right in Fig.~\ref{fig:coreform} has initial conditions very similar to those of Pluto: size bin 1200~km and $X_s=0.80$. The surface is hotter and the mass is smaller by about 6\%, but these are close enough to warrant a comparison with the present stratified structure inferred from observations \citep{Spencer20}. In our model, core formation starts early, both due to the high rock content, and due to the relatively low central pressure and hence melting temperature, compared to the other objects. As explained above, oscillations on the thermal time scale of the core, $\sim10^9$~yr, are due to the effect of water flooding the ice-depleted core, refreezing and melting at a later stage. 

\begin{figure}[h!]
\centering
\includegraphics[scale=0.3]{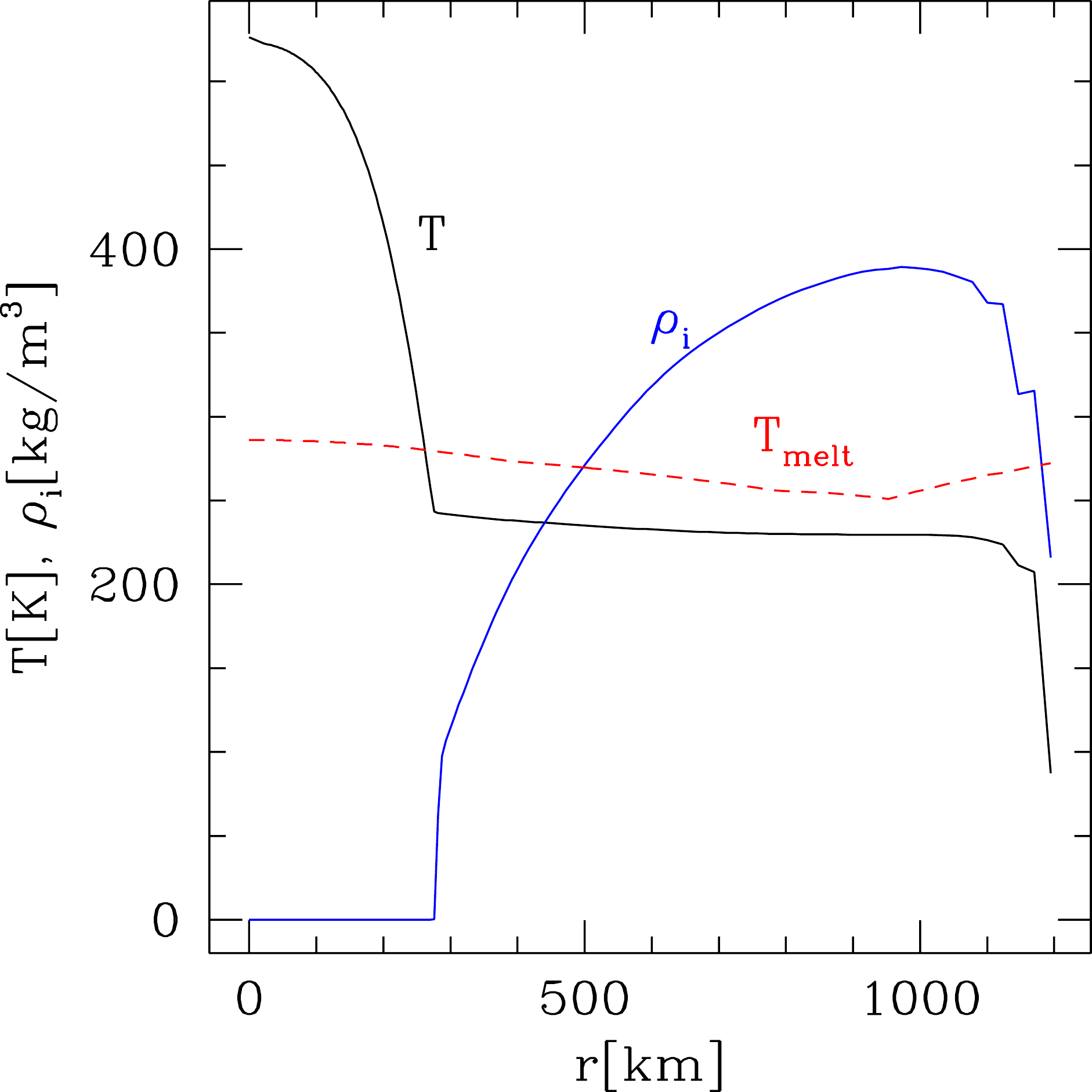}
\caption{\small{Present day internal structure for the model of  1200~km and $X_s=0.80$: temperature and H$_2$O density ($X_i\rho$) profiles, and the melting temperature profile. Only a small amount of heat would be needed to form a subsurface ocean between $\sim500$ and 1000~km.}}
\label{fig:Pluto-ocean}
\end{figure}
The present day internal structure of this model is shown in Fig.\ref{fig:Pluto-ocean}. We note that the temperature throughout the ice mantle is almost constant and only slightly below the local melting point. The difference is smaller than the effect that would result from the presence of impurities, such as ammonia, which were neglected in the present study. Thus it is quite possible that this layer be liquid rather than icy. Alternatively, a small amount of additional heat provided by tidal forces exerted by Charon may turn the ice layer into an internal ocean \citep{Saxena18}, although the effect may be ambiguous \citep{Robuchon11}. We now turn to the present-day structure of all models.

\newpage

\subsection{Present internal structure}
\label{ss:present}

\begin{figure}[hp!]
\centering
\includegraphics{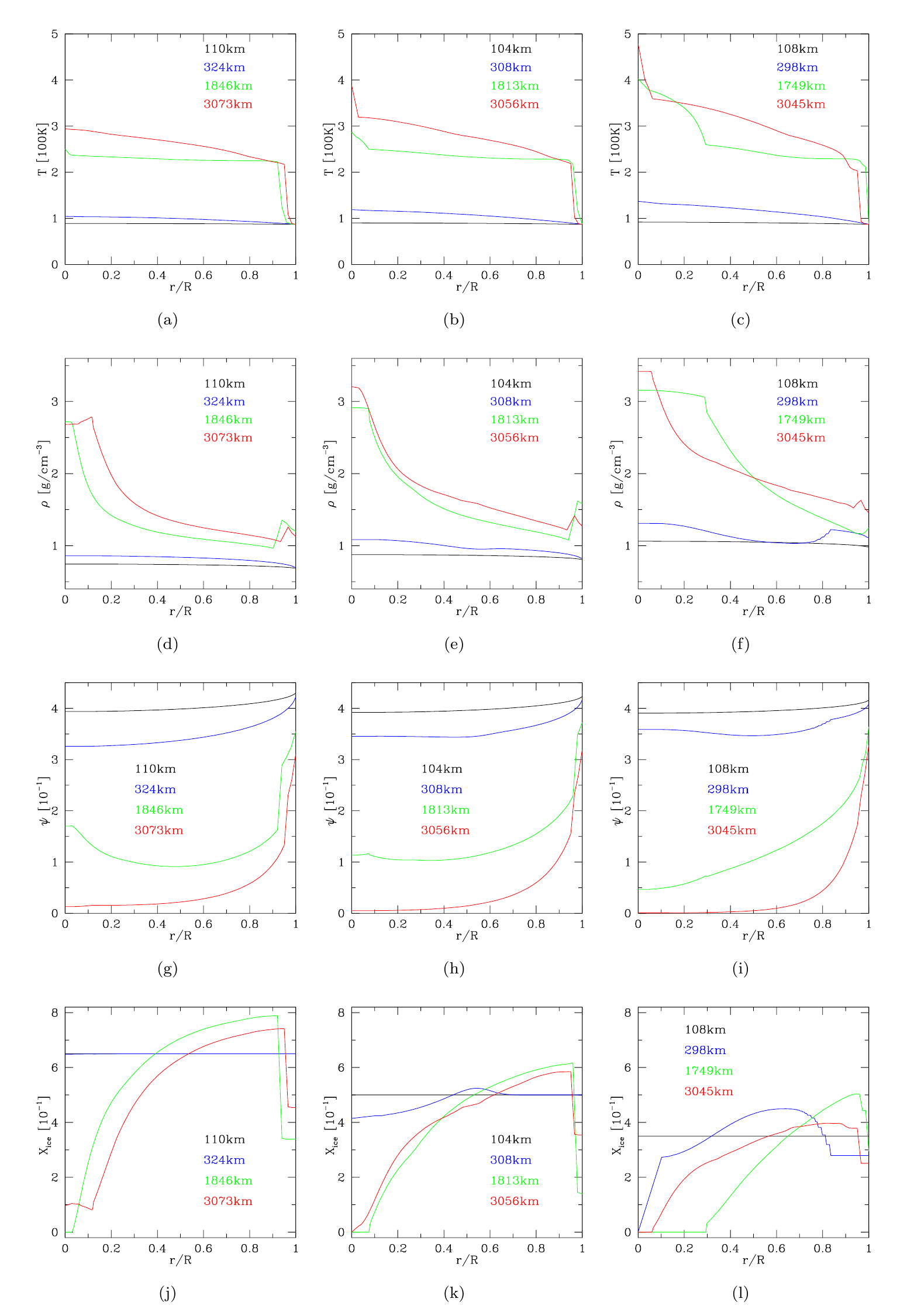}
\caption{\small{Profiles of the temperature (top row panels),  density (second row panels), porosity (third row panels) and mass fraction of ice (bottom panels) as function of normalized distance from the center, at the end of evolution, for representative models (marked by their final radii): initial rock mass fraction of 0.35 (left column), 0.50 (middle column) and 0.65 (right column).}}
\label{fig:structure}
\end{figure}

The general structure that emerges for most objects at the end of evolution, after 4.5~Gyr, is stratified, comprised of a rocky core, an ice-rich mantle---with an ice/rock ratio exceeding the initial bulk value---and an outer dense crust. Features of the present structure are given in Fig.~\ref{fig:structure}. Each row shows a different property---temperature, density, porosity and ice mass fraction---and each column represents a different initial composition. The main features may be summarized as follows. The internal temperatures increase with increasing rock content and with increasing mass.
Only small ($R\le 300$~km) ice-rich ($X_i/X_s>1$) objects retain a porous, almost homogeneous structure. Very large objects have a very low porosity almost throughout, except in a relatively thin outer layer. The ice-depleted core is clearly seen in the density profiles (not only in the ice density profiles); we note that differentiation is gradual rather than sharp. 

\begin{figure}[h]
\centering
\includegraphics{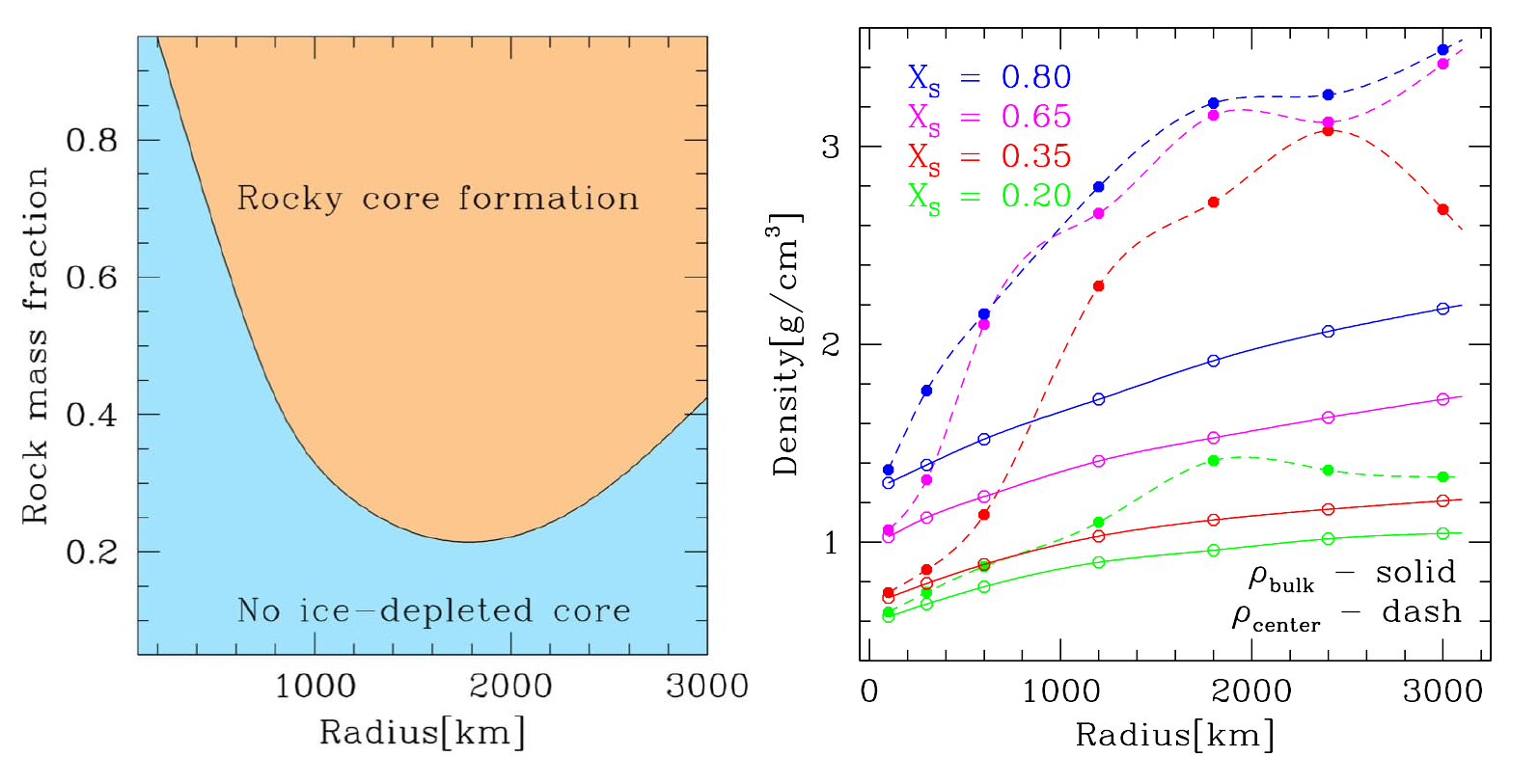}
\caption{\small{Left:  Rough division of the initial parameter space between the region where an ice-depleted core forms and the region where there is no complete differentiation. The effect of high melting temperatures at very high pressures is clearly illustrated. Right: Bulk (average) densities and central densities of a sample of models at the end of evolution. The central density increases with size (mass) and with initial rock content, but the trend changes when no core is formed. The models that did not form an ice-depleted core are all those with radii of 100~km or less and all those with an ice content of 80\% by mass. In addition, among those shown: $X_s=0.35$ and $R$=300, 600 and 3000~km.}}
\label{fig:icetorock}
\end{figure}

Differentiation is naturally a function of rock content, but it is also a function of size. For large (massive) objects, which reach high pressures in the central part, the melting temperature there is also very high, and if the rock content is not high enough to provide sufficient radioactive heat, the temperature remains below the melting point and full differentiation does not occur. The location in parameter space where rocky cores are formed is shown in left-hand panel of Fig.~\ref{fig:icetorock}. The presence of a core is also reflected by the higher central densities of the respective objects, as shown in the right-hand panel of Fig.~\ref{fig:icetorock}. The contrast between bulk and central density shown in the figure illustrates the tendency to central condensation as objects are larger, compared with almost constant-density small objects.

We note that $X_s=0.35$ intersects the dividing boundary of core formation twice, at small and at large radii. This explains the apparently strange variation of central density  with radius for this case: the larger body, of 3000~km radius has a lower central density than the smaller one, with $R=2400$~km, although one would expect it to be more compressed. Indeed, it is more compressed---the central pressure is 2.6GPa and the porosity is 0.013, as compared to 1.4GPa and 0.067, respectively, for the smaller body---but since it does not form a rocky core, the large proportion of ice leads to a lower density (see Fig.~\ref{fig:eost}).

In most objects, the ice content drops in a subsurface layer. This is because the water and vapor, flowing outwards from the hot interior and encountering a steep temperature gradient, freeze before reaching the surface, forming an icy barrier. In many cases, this results in a local density peak; at the same time, since the dominant component is rock, which has a high specific density, the porosity is higher. We regard this layer as a crust; its thickness is on the order of tens of km. The same effect of high density together with relatively high porosity is also seen in the rocky cores.

As a result of serpentinization and dehydration, the rock structure may be found in three different states: unaltered, if temperatures remained below serpentinization temperature at all times; hydrated, where temperatures have exceeded this temperature while water was still present; and dehydrated, where temperatures reached sufficiently high levels for water to be driven out of the rock. Objects of different radii have different distributions of these types of rock. 
\begin{figure}[h!]
\centering
\includegraphics[scale=0.38]{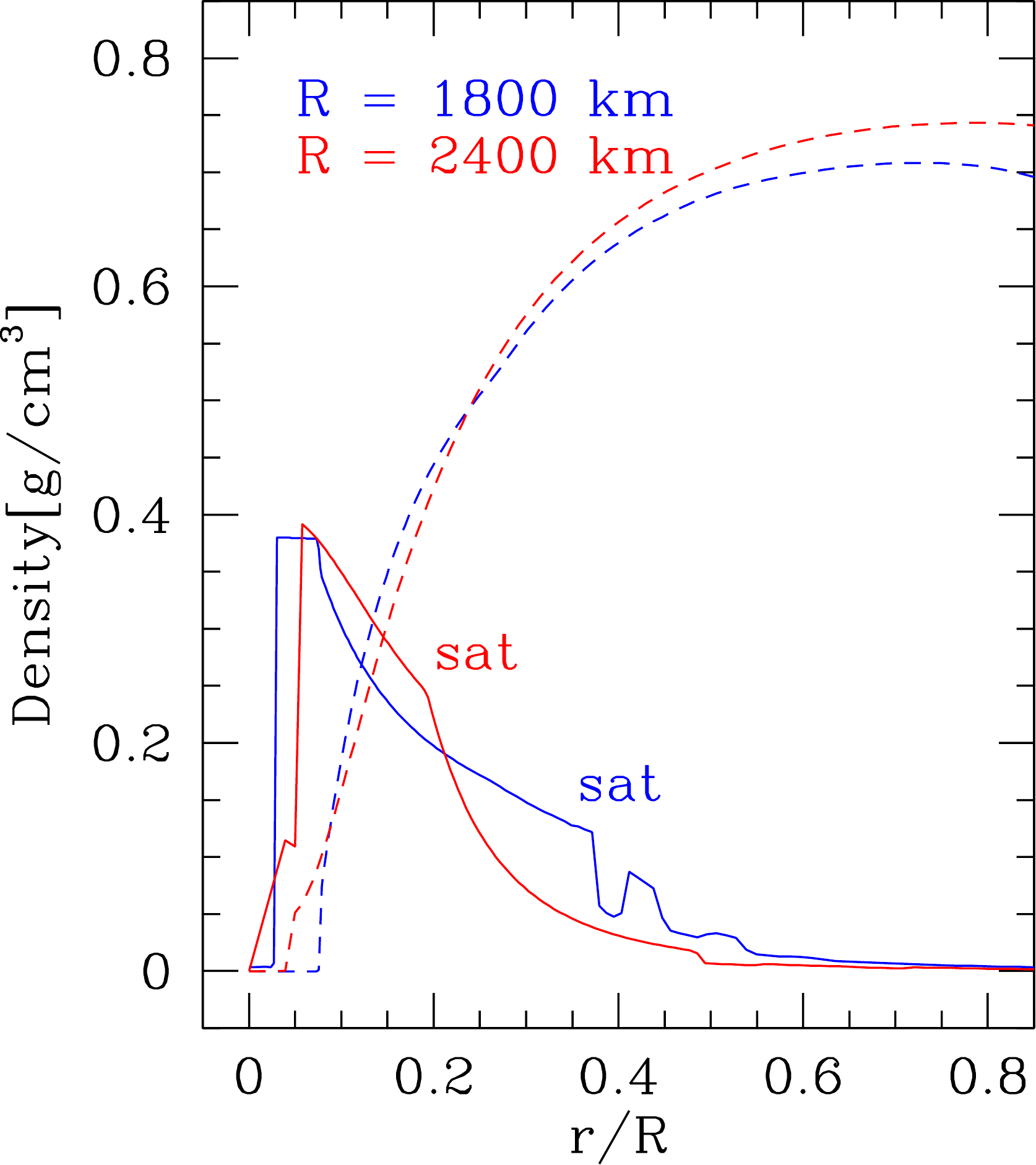}
\caption{\small{Profiles of the mass of water per unit volume contained in the hydrated rock (solid lines), for models of the same initial rock content of 0.50 by mass, but different initial radii, as indicated. The partial density of free ice is also shown (dashed lines) for the same models. The boundary of the saturated hydrated rock is marked (sat).}}
\label{fig:serp}
\end{figure}

To illustrate this effect, we show in Fig.~\ref{fig:serp} two examples of profiles of the mass of water per unit volume that is absorbed in the hydrated rock, for objects of different radii, but the same initial fraction of rock, $X_s=0.50$. The rock in the cores of these objects is hydrated up to saturation; the densities differ because the total rock densities differ. The same occurs down to radii of $\sim600$~km. The rock outside the core is largely unaltered (or only slightly hydrated), since temperatures in these regions never rise to serpentinization values. The largest and most rock-rich bodies, which reached temperatures above 700~K, so that the cores underwent dehydration, contain all three types of rock: a mostly dehydrated-rock core, a partly hydrated-rock layer and unaltered rock above it. These outer layers also contain water ice. 

\subsection{Effect of advection}

A crucial factor that affects the course of evolution is heat conduction by advection, which constitutes a special feature of the model presented here that is not included in other models of icy bodies (although it is a common feature in most comet models). Advection accompanied by phase transitions has long been considered as an important means of heat conduction \citep{Steiner91}. We find that it may assume a wide range of different behaviors: from being negligible compared to heat conduction, to being dominant; from enhancing conduction in the same direction, to opposing it, because mass flows are controlled by composition gradients on top of the temperature gradient, and these may have opposite signs.
\begin{figure}[h!]
\centering
\includegraphics[scale=0.32]{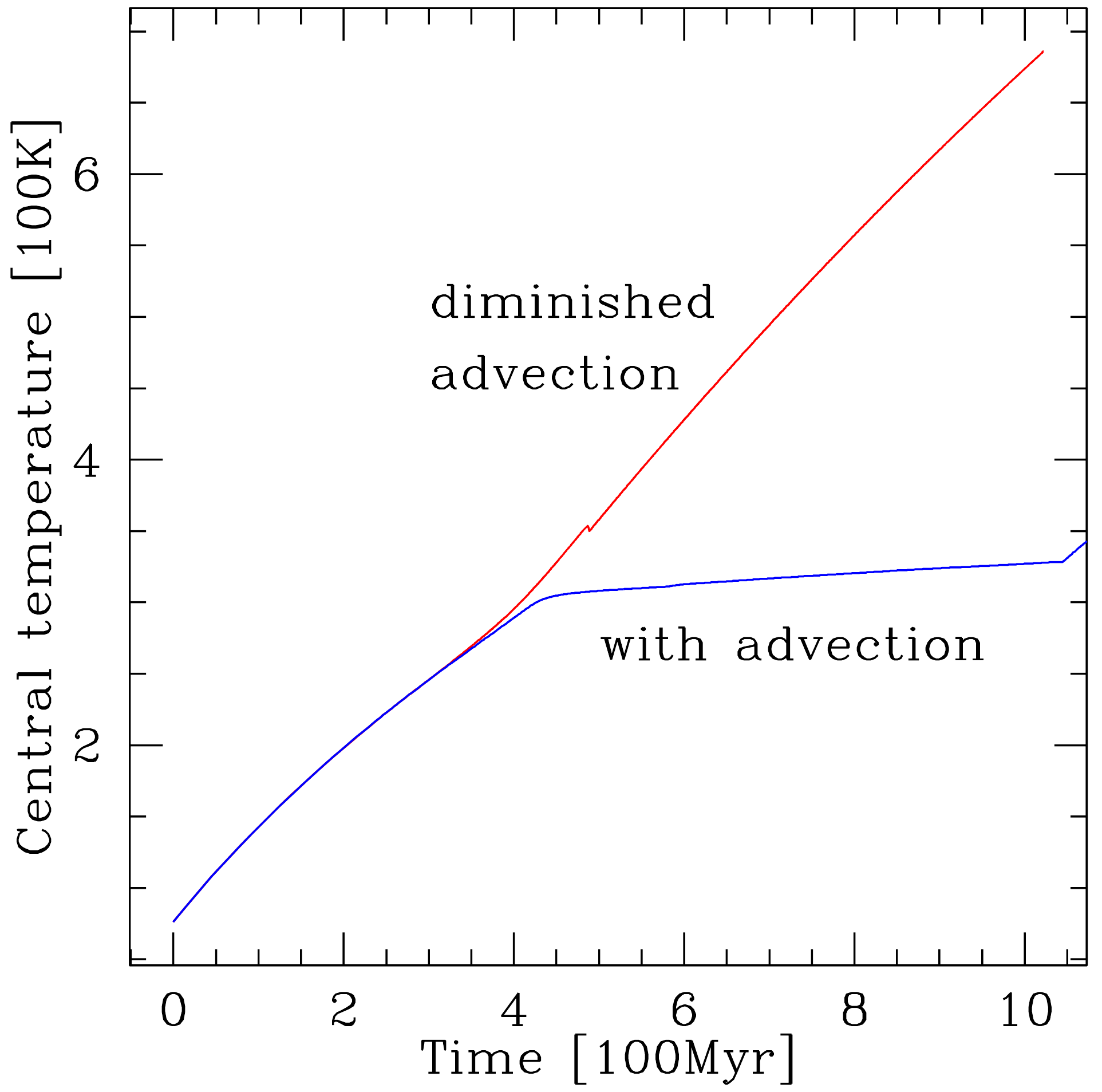}
\caption{\small{Comparison between evolutionary runs with advection and with diminished advection by a uniform factor of 0.3, showing the central temperature of models with radii of 2400~km and $X_s=0.65$.}}
\label{fig:advec}
\end{figure}

Our models reach lower internal temperatures than those obtained for example for Pluto models (radius of 1200~km) that assume an initially differentiated structure into a silicate core and icy mantle \citep[e.g.,][]{Robuchon11, Bierson18}. Although the approach and assumptions are different, it is instructive to test the effect of advection on the heat flow. We have therefore run a comparison model for $R=2400$~km and $X_s=0.65$ for $10^9$~yr, reducing the advective flux by a factor of 0.3 (that is, 70\%) everywhere. The results for the evolution of the central temperature are shown in Fig.\ref{fig:advec}; they clearly show that advection by flowing vapor and water strongly affects (impedes) the rise in temperature in the central part of the body, which starts much earlier in the comparison model.

\newpage

\subsection{Effect of boundary condition}

\begin{figure}[h!]
\centering
\includegraphics[scale=0.4]{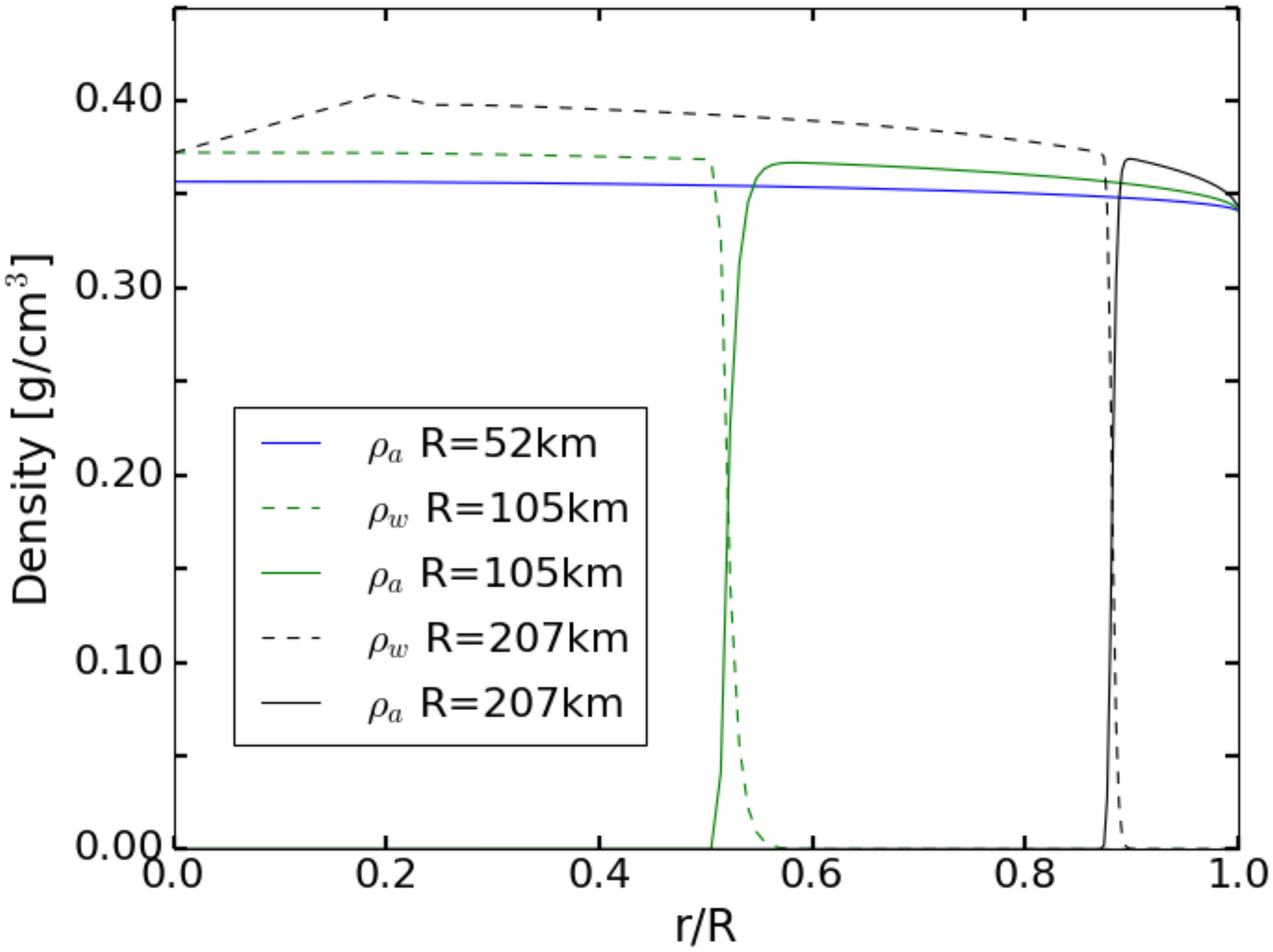}
\caption{Profiles of the ice---amorphous and crystalline---partial densities at the end of evolution for small bodies; symbols are $\rho_a=X_a\rho$ and $\rho_w=X_c\rho$. 
}
\label{fig:distant2}
\end{figure}
We have computed a series of models for a very low equilibrium temperature (20~K) and an initial rock mass fraction $X_s=0.65$, assuming the ice to be in amorphous form, which implies long formation times and hence a negligible effect of early heating by short-lived radionuclides. Similarly to the previous models, the internal temperatures start rising; the ice crystallizes releasing more heat, and in most cases the crystallization front sweeps the entire object, except for an outermost layer that preserves the amorphous ice. 

Small bodies, up to about 100~km in radius are colder throughout compared to the less distant counterparts and this is because being smaller, their surface to volume ratio is larger and they cool more efficiently. They do not reach sufficiently high temperatures for crystallization to occur and hence they retain the amorphous ice throughout, as shown in Fig.~\ref{fig:distant2}. In all objects, the outer part remains very cold and the ice is preserved in amorphous form down to tens of km below the surface for bodies of a few hundred km in radius and up to 200~km for the 3000~km body. The amorphous ice may contain significant amounts of volatiles trapped in it. In fact, the outer, cold layer could retain other volatiles also as ices; close to the surface, even extremely volatile species such as CO and CH$_4$ may survive. 

Perhaps counterintuitively, the large distant objects reach slightly higher internal temperatures, when compared with the closer one of the same size and ice/rock mass ratio. The reason is twofold: first less energy is emitted, as the surface temperature rises only slightly above the equilibrium temperature and secondly, there is an additional heat source due to the crystallization of the initially amorphous rather than crystalline ice. The small objects approach equilibrium with the environment, hence the more distant ones have colder interiors.

\subsection{General relations}

\begin{figure}[h!]
\centering
\includegraphics[scale=0.40]{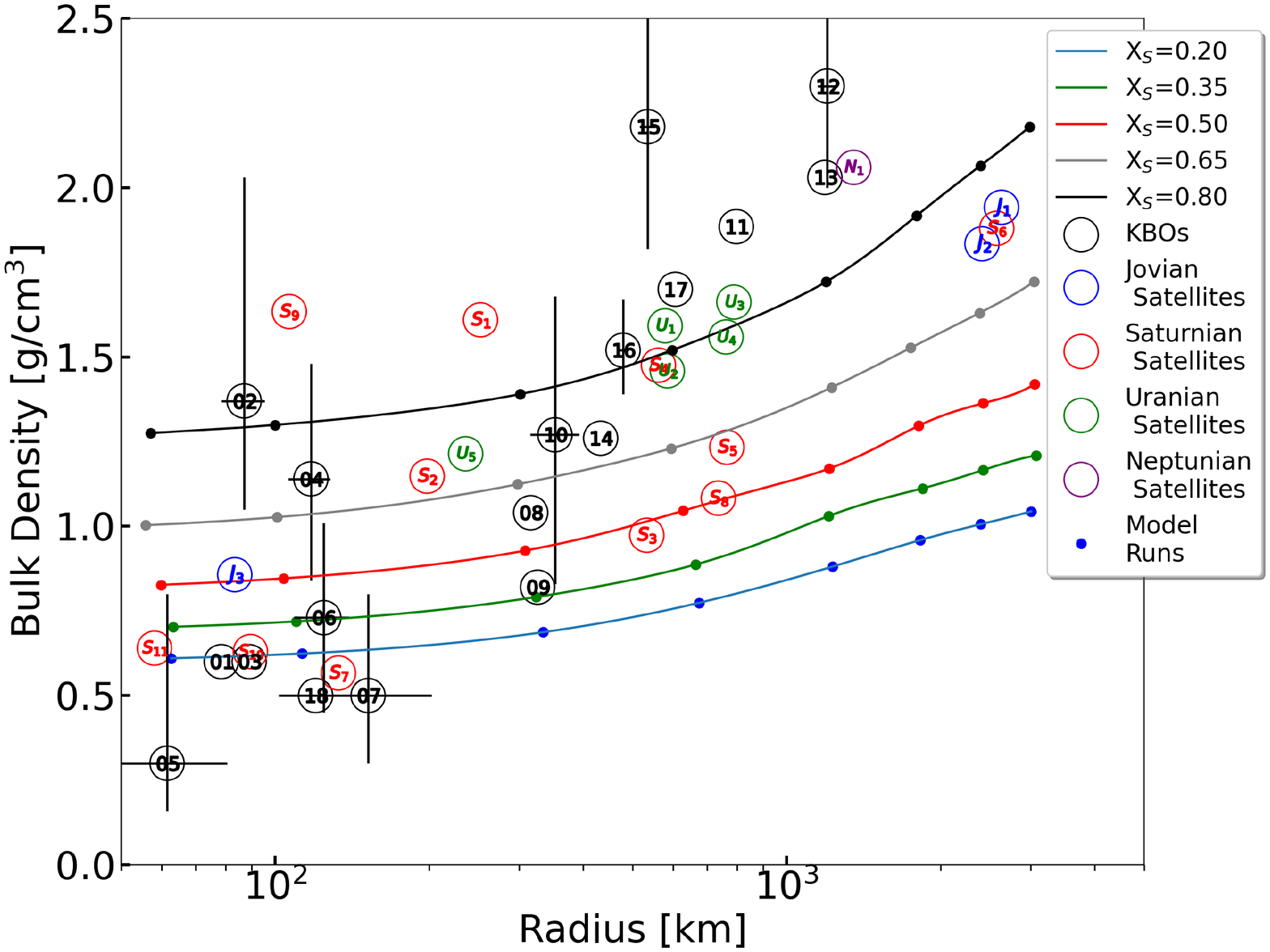}
\caption{\small{Bulk density -- total mass divided by total volume -- as function of radius at the end of evolution for all models. Observational results -- listed in Tables~\ref{tab:kbos} and \ref{tab:sat} -- for Kuiper belt objects and planetary satellites are marked for comparison.}}
\label{fig:RhoofR}
\end{figure}

In order to test our results by observations, we show in  Fig.~\ref{fig:RhoofR} the  correlation between bulk (average) density and radius, for which there are estimates for various objects in our solar system, such as satellites and KBOs, as listed in Tables~\ref{tab:kbos} and \ref{tab:sat}.
The differences in bulk density that are clearly seen in the figure may result from two independent factors---porosity and composition---and by observation alone it is not possible to determine either. Models, however, are able to determine both. Our results show that most objects in the samples
are found between the curves corresponding to $X_s=0.65$ and 0.8 (rock/ice mass ratio $\sim$2-4). For each composition and size, the porosity profile within a body may be obtained, as shown by the examples of Fig.~\ref{fig:structure}. In principle---keeping in mind simplifications and uncertainties---we should also be able to predict whether or not a given object possesses a rocky core by cross-correlating the results presented in Figs.~\ref{fig:RhoofR} and \ref{fig:icetorock}. For example, our results indicate that Enceladus ($S_1$), with a very high rock content ($X_s > 0.8$), should be differentiated despite its small size.
\begin{deluxetable*}{llccl} [h!]
\tablenum{3}
\tablecaption{Radii and densities obtained from KBO observations}
\tablewidth{0pt}
\tablehead{
\colhead{Fig.\ref{fig:RhoofR}} & \colhead{Name} &  \colhead{Radius [km]} &  \colhead{$\rho_b$ [g cm$^{-3}$] } &  \colhead{Source} 
}
\label{tab:kbos}
\startdata
 1 & Typhon & 78.5 & 0.6 & \cite{Typhondata} \\
 2 & Ceto & 87 & 1.37 & \cite{Cetodata} \\ 
 3 & Teharonhiawako & 89 & 0.6 & \cite{88611_Altjira_Sila_Vardadata} \\ 
 4 & 88611 2001 QC 298 & 117.5 & 1.14 & \cite{88611_Altjira_Sila_Vardadata} \\
 5 & Altjira & 61.5 & 0.3 & \cite{88611_Altjira_Sila_Vardadata} \\
 6 & Sila & 124.5 & 0.73 & \cite{88611_Altjira_Sila_Vardadata} \\
 7 & Lempo & 152 & 0.5 & \cite{Lempodata} \\
 8 & 229762 2007 UK 126 & 316 & 1.04 & \cite{229762_2007data} \\
  9 & 55637 2002 UX 25 & 326 & 0.82 & \cite{55637_2002_data} \\
 10 & Varda & 352.5 & 1.27 & \cite{88611_Altjira_Sila_Vardadata} \\
 11 & Haumea & 797.5 & 1.885 & \cite{Haumea_data} \\
 12 & Eris & 1200 & 2.3 & \cite{Eris_data} \\
 13 & Pluto & 1188 & 1.85 & \cite{Pluto_and_Charon_Data}\\
 14 & Salacia & 433 & 1.26 & \cite{Salacia_data}\\
 15 & Quaoar & 535 & 2.18 &  \cite{88611_Altjira_Sila_Vardadata} \\
 16 & Orcus & 479 & 1.52 & \cite{TNOsRcool_Orcus} \\ 
 17 & Charon & 606 & 1.7 & \cite{Pluto_and_Charon_Data} \\ 
 18 & Pholus & $\sim$120 & 0.5 & \cite{Pholus_data} \\
 \enddata
 \end{deluxetable*}
 \begin{deluxetable*}{llccl} [h!]
\tablenum{4}
\tablecaption{Radii and densities obtained from observations for satellites of Jupiter, Saturn, Uranus and Neptune}
\tablewidth{0pt}
\tablehead{
\colhead{Fig.\ref{fig:RhoofR}} & \colhead{Name} &  \colhead{Radius [km]} &  \colhead{$\rho_b$ [g cm$^{-3}$] } &  \colhead{Source} 
}
\label{tab:sat}
\startdata
 $J_1$ & Ganymede & 2631.2 & 1.942 & \cite{Ganymede_data}\\ 
 $J_2$ & Callisto & 2410.3 & 1.834 & \cite{Callisto_data} \\ 
 $J_3$ & Amalthea & 83.45 & 0.849 & \cite{Amalthea_data} \\
 \hline
  $S_1$ & Enceladus & 252.1 & 1.6096 & \cite{Enceladus-Phoebe_data} \\ 
 $S_2$ & Mimas & 198.3 & 1.15 & \cite{Enceladus-Phoebe_data} \\ 
 $S_3$ & Tethys & 533 & 0.97 & \cite{Enceladus-Phoebe_data}\\ 
 $S_4$ & Dione & 561.7 & 1.48 & \cite{Enceladus-Phoebe_data}\\ 
 $S_5$ & Rhea & 764.3 & 1.23 & \cite{Enceladus-Phoebe_data}\\ 
 $S_6$ & Titan & 2575.5 & 1.88 & \cite{Enceladus-Phoebe_data}\\ 
  $S_7$ & Hyperion & 133 & 0.57 & \cite{Enceladus-Phoebe_data} \\ 
  $S_8$ & Iapetus & 735.6 & 1.083 & \cite{Enceladus-Phoebe_data} \\ 
  $S_9$ & Phoebe & 106.6 & 1.63 & \cite{Enceladus-Phoebe_data}\\ 
  $S_{10}$ & Janus & 89.5 & 0.63 & \cite{Janus-Epimetheus_data} \\ 
  $S_{11}$ & Epimetheus & 58.1 & 0.64 & \cite{Janus-Epimetheus_data} \\ 
 \hline
 $U_1$ & Ariel & 578.9 & 1.59 & \cite{Uranus-GM_data}, \cite{Uranus-R_data} \\ 
 $U_2$ & Umbriel & 584.7 & 1.46 & \cite{Uranus-GM_data}, \cite{Uranus-R_data}\\ 
 $U_3$ & Titania & 788.9 & 1.66 & \cite{Uranus-GM_data}, \cite{Uranus-R_data}\\ 
 $U_4$ & Oberon & 761.4 & 1.56 & \cite{Uranus-GM_data}, \cite{Uranus-R_data}\\ 
 $U_5$ & Miranda & 235.8 & 1.21 & \cite{Uranus-GM_data}, \cite{Uranus-R_data}\\ 
 \hline
$N_1$ & Triton & 1353 & 2.06 & \cite{Triton_data} \\
\enddata
\tablecomments{The mean densities for the Uranian Satellites were taken from the NASA website: https://ssd.jpl.nasa.gov, which uses the $GM$ and $R$ data from \cite{Uranus-GM_data} and \cite{Uranus-R_data}.
}
\end{deluxetable*}

\section{Discussion}
\label{s:discussion}

The purpose of the present study was to investigate the redistribution of water ice in the course of long-term evolution for small to intermediate-size bodies in general, considering initial ice to rock mass ratios in the range 0.25-4 and a range of radii between 50-3000~km. The upper size limit was chosen so that the porosity at the center would not completely vanish and internal temperatures would remain below the melting temperature of rock. The ambient temperature where the bodies evolved was chosen to be low enough to prevent ice loss by sublimation at and near the surface, so as to preserve the ice/rock ratio. The same conceptual model was used in all cases, namely a porous rock matrix with porous ice embedded in it, and the same initial conditions of cold homogeneous bodies, implying that the energy deposited by the accretion process was not sufficient for melting and early differentiation. We did not attempt to model any particular object and were not biased in the choice of initial conditions by any observational result. Rather, our goal was to provide a theoretical basis for the general understanding of the ice-rock differentiation phenomenon.

Compression due to self gravity, heating by radioactive decay, migration of water and vapor through pores, and rock-ice interactions were taken into account and found to change the structure of most of these bodies considerably during 4.5~Gyr of evolution. 
Upon sufficient heating, the pores in the interior become filled with vapor, and eventually water, which migrate to the cold outer layers, leaving behind a rocky core and building up an ice-rich mantle. 

Only H$_2$O ice was included in the model, although other ice species must have been present in the disk where the bodies formed. Their abundances, however, would be much lower than that of H$_2$O and their effect on the internal structure of the bodies considered should be much smaller. These ices would evaporate in the interior, while H$_2$O would still be in the ice phase, and they would flow towards the cold outer layers. The species with evaporation temperature above the surface temperature would refreeze in the outer layers, which would thus be composed of a mixture of ices. The more volatile species would be lost (or form an atmosphere). These processes would consume a small fraction of the released energy and hence would not affect the differentiation between water and rock. They would have a marked effect on the atmosphere of the body, if such existed. The atmosphere itself only affects a thin outer  layer rather than the deep interior, where ice melting and differentiation takes place. Hence both trace elements and the possible presence of an atmosphere have been ignored. They are  beyond the scope of the present study, whose main objective is ice-rock differentiation. 

Differentiation between ice and rock may occur in two ways, both of which invoke melting of the ice upon heating: either the liquid diffuses outwards, leaving behind a rocky core that becomes compressed by self-gravity---the model adopted in the present study---or, rock particles settle to the center through the liquid, displacing it outwards to form an icy mantle \citep[e.g.,][and references therein]{Neumann19}. The end results are similar, but in the former model, the icy mantle still contains rock, while in the latter, it is pure ice. The diffusion of both vapor and water is controlled by the porosity, by the pore sizes and the distribution of pores. There are almost endless possibilities of adjusting constitutive relations for these properties, when so few constraints are available. We have adopted a simple set of such relations and applied it to all the combinations of initial parameters, in order to single out the effect of initial mass and rock/ice ratio and discern trends of behavior. We ran tests with different physical properties, but these did not alter our main qualitative conclusions regarding differentiation by the mechanism adopted in this study.

Our main conclusions may be summarized as follows:
\begin{itemize}
\item{}Small bodies of radii up to a couple of hundred km retain a homogeneous composition and stay cold throughout evolution. In larger bodies, a rocky core begins forming, depending on the ice/rock mass fraction. The cores are generally smaller than those inferred phenomenologically for the few well-observed bodies. The core size depends however on the physical parameters assumed (such as permeability and conductivity), which are still uncertain. 
The transition from the rocky core to the ice-rich outer region is generally gradual. There is no marked boundary between the two regions.
Small to intermediate objects retain a relatively high porosity in the core, even when differentiation occurs. This conclusion may be relevant to the explanation of the activity of Enceladus \citep{Choblet17} that invokes water flow through a porous core.
\item{}When the ice melts and the temperature rises, the rock in the core becomes hydrated by serpentinization, absorbing some of the initial ice. However, bodies with radii $\sim3000$~km and a high initial rock content reach high enough temperatures in the core to become dehydrated. Both processes compete with water flow and lead to oscillations in the core size until the water is completely removed, or permanently retained by the rock (when temperatures drop below dehydration limit).
The dependence of melting temperature on pressure above $\sim10$~MPa, in particular the steep rise at GPa pressures, has a marked effect on the evolution: in big (massive) objects it delays core formation to $\sim1$~Gyr and prevents core formation altogether in large objects that are ice-rich (low rock content). This effect, too, may cause the core size to oscillate, as water produced in outer layers after core formation may migrate into the core and refreeze despite the high temperature prevailing there. This is because the pressure in the core increases in the interim, and drives the melting temperature still higher.
 \item{}The ice-rich mantle below the surface comprises a large fraction of the volume in all models. If objects are captured by planets, this mantle may be activated by the additional energy sources that become available, such as tidal heating. In this case a sub-surface ocean can form and allow for cryovolcanism  \citep{rhoden2015, thomas2016, 
 Guilbert-Lepoutre20}. 
A dense outer crust is found in differentiated objects. It results from the steep temperature gradient near the surface, which blocks sharply the flow of liquid water or vapor from the interior, causing rapid freezing. Interestingly, the same effect, although in reverse, has been found in simulation models of comets, both theoretical and experimental: in that case, the sharp temperature gradient at the surface goes in the opposite direction (from the hot surface towards the cold interior) and vapor resulting from sublimation in the surface layer that flows inwards freezes forming a dense sub-surface crust \citep{Prialnik91,Spohn89}.  Very close to the surface, the relatively ice-poor layer is porous even in the largest bodies. Internal stresses caused by changes in density and pressure may show on the surface as fissures or cracks. A dense, undifferentiated crust of $\sim$100~km typical thickness was also found in evolutionary studies of KBOs in the range 100-1000~km by \cite{Desch09}, based on the rock-settling differentiaion scenario. \cite{Rubin14} studied the effect of Rayleigh-Taylor instabilities in such a layer and concluded that a crust of $\sim$60~km should persist due to the high viscosity of ice below about 150~K, which would prevent the overturn of the crust.
\item{}Testing the effect of a much lower ambient temperature ($\sim$20~K), and hence starting with the water ice in amorphous state, we found the general trends of evolution to be similar, except that the central temperatures attained in large objects are somewhat higher, while the outermost layers remain cold enough to preserve the amorphous ice. Bodies of only a few tens of km in size retain the amorphous ice throughout. 
\end{itemize}

\noindent Observations of bodies of the sizes considered here---KBOs, icy satellites and small exoplanets---yield estimates for radii and masses, from which the dependence of bulk density on radius may be obtained. Our results show that generally, bulk densities---which vary between 0.6 and 2.2 g~cm$^{-3}$---increase with increasing radius and with increasing rock content. The observed objects, with only a few exceptions, are found in the region where the rock mass fraction exceeds 0.5, most of them in the range $X_s=0.65-0.80$. The same general trends were obtained by  \cite{Bierson19} in a theoretical study of the effect of porosity on the mass distribution in Kuiper belt objects, adopting a different model than that presented here, in which melting and differentiation were not taken into account. The difference between the results of these two studies lies in the shape of the variation of the bulk density with radius for a given initial composition, which has a marked flexing point in Bierson \& Nimmo's results, while it is smooth according to our model. We have argued that knowing the mass and radius of a body should enable the determination of the ice/rock mass fraction and the prediction of the presence or absence of a rocky core.

The main conclusion of our systematic parameter study is that both initial mass and initial composition have crucial effects on the evolution of an object and the resulting present structure: for the same ice/rock ratio, objects of even slightly different masses may have completely different behaviors, 
and similarly, keeping the mass constant but slightly varying the ice/rock ratio may lead to radically different configurations. This diversity of evolutionary outcomes will very likely persist even when physical parameters are improved or updated.  Models are simplified versions of reality and they can always be expanded and improved to simulate the real world ever more accurately. Observations will guide us in our next model upgrades.
\\
\\
\noindent {\textbf {Acknowledgements}} We acknowledge with thanks the support of the Israeli Science Foundation for this research through the ISF grant 566/17. We thank an anonymous referee for a careful reading of the original manuscript and for very useful comments and suggestions.

\newpage

\bibliographystyle{aasjournal}
\bibliography{ApJbib}

\end{document}